\documentclass[aps,pre,onecolumn,showpacs,showkeys,a4paper]{revtex4}
\pdfoutput=1
\input epsf  
\usepackage{graphicx} 
\usepackage{amsmath}
\usepackage{amssymb}
\usepackage{amscd}

\begin{document} 

 \title{Some Exact Results for the Exclusion Process}
\author{Kirone Mallick}
  \affiliation{Institut  de Physique Th\'eorique, C. E. A.  Saclay,
 91191 Gif-sur-Yvette Cedex, France}
\pacs{05-40.-a;05-60.-a }
 \keywords{Non-equilibrium statistical physics; ASEP; Bethe Ansatz; large deviations, Matrix Ansatz.}
  \date{24th of November 2010}

 \begin{abstract} 
 The  asymmetric simple  exclusion process (ASEP) is a paradigm for
 non-equilibrium physics that appears as a building block to model
 various  low-dimensional transport phenomena, ranging  from
 intracellular  traffic  to quantum dots.  We review some recent
 results obtained for the system  on a  periodic ring by using the
 Bethe Ansatz. We show that this method allows to derive analytically
 many  properties of  the dynamics of the model such as  the spectral
 gap and  the  generating  function  of the current.  We also
 discuss  the solution of a generalized  exclusion process   with
 $N$-species of particles and explain how a geometric construction
 inspired from queuing theory sheds light on the Matrix Product
 Representation technique that has been very  fruitful to derive exact
 results for the ASEP.
  \end{abstract}

\maketitle 

 \section{Introduction}

 Equilibrium Statistical Mechanics is a well-established field  that
 embodies Classical Thermodynamics (see e.g. F. Reif,  1965 \cite{Reif}). In an  extremely
 terse formulation,  this  theory could be summarized by saying that
 the  relevant features of the  microscopic physics of  any  system
 are   encoded in a  probability measure for which, at  thermal equilibrium,  a  mathematical
 formula  is known. Indeed,  a  system in
 contact with a heat bath at temperature $T$  occupies all  accessible
 microscopic configurations ${\mathcal C}$  with a probability
 $P_{{\rm eq}}({\mathcal C})$, given by the  Boltzmann-Gibbs canonical
 law:
\begin{equation}
{ P_{{\rm eq}}({\mathcal C})  =
  \frac{{\rm e}^{ - E({\mathcal C}) /kT}}{Z}}\,,
  \label{Canonical}
\end{equation}
 $E({\mathcal C})$ being  the energy of the configuration ${\mathcal
 C}$.  The  Partition Function $Z$  (or `sum over states') is related
 to the thermodynamic   Free Energy  F via the relation
\begin{equation}
  {  F = -kT {\rm Log}\,Z  } \,. \nonumber
 \end{equation} 
  This relation is  nothing but  an avatar of the celebrated Boltzmann
  law, $S = k\,{\rm Log}\,\Omega$.  Macroscopic  observables are
  obtained by taking the expectation values of the corresponding
  microscopic observables   with respect to  the  canonical
  measure~\eqref{Canonical}.   These postulates  provide us with a
  {\it well-defined prescription}  to analyze  systems  {\it at
  equilibrium}. In particular,  equilibrium  statistical mechanics
  predicts  macroscopic  fluctuations  (typically Gaussian) that are
  out of reach of Classical  Thermodynamics:   the  paradigm of  such
  fluctuations is the Brownian Motion.

\begin{figure}[ht]
\begin{center}
  \includegraphics[height=2.0cm]{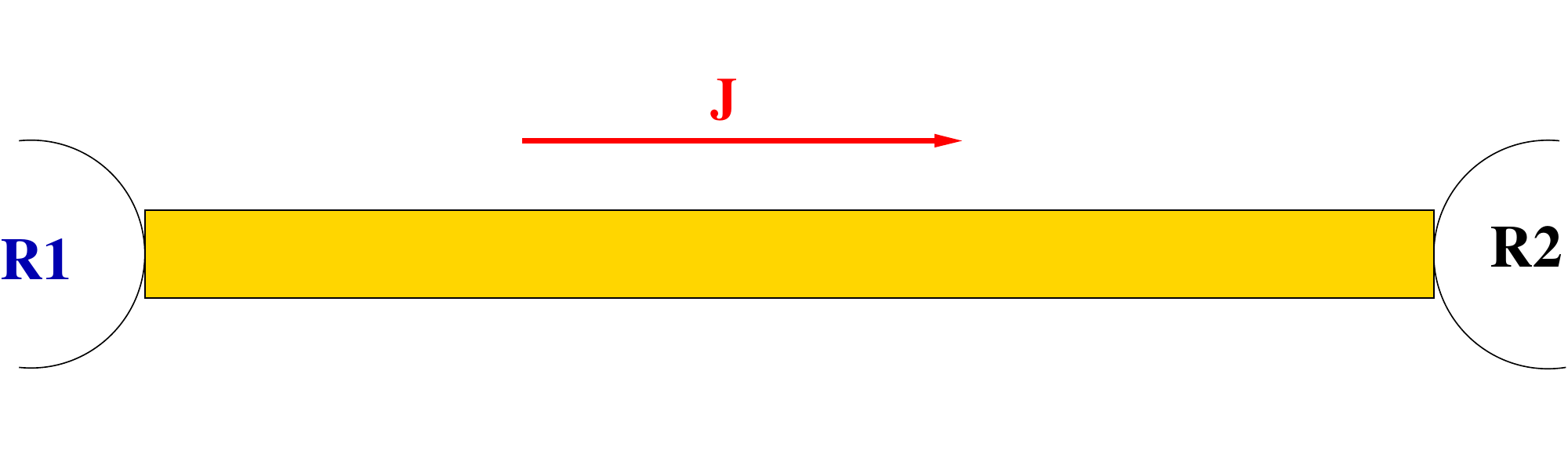}
  \caption{A  stationary driven system  in contact with 
 two reservoirs at different temperature and/or potential.}
\label{fig:courant}
 \end{center}
\end{figure}

   For systems far from equilibrium, a  fundamental theory that would
   generalize  the formalism of
   equilibrium  Statistical Mechanics  to  time-dependent processes 
   is  not yet available. A
   schematic non-equilibrium process  can be represented as a  rod  in
   contact with  two reservoirs at different temperatures, or  at
   different electrical (chemical) potentials (see
   Figure~\ref{fig:courant}).  In the  stationary regime, a constant
   current flows through  the system. Many  fundamental questions
   remain   to be answered in order to understand the  physics  of
   such a simple system:    What are  the  relevant  parameters that
   would fully  characterize  the macroscopic state  of a
   non-equilibrium system?  Can the stationary state be defined as the
   optimum of some  (unknown)  macroscopic functions of some
   (unknown)  parameters?  Does a  general equation of state exist?
   Can one classify non-equilibrium processes into  `Universality
   classes'? Can one postulate a  general form for   some
   non-equilibrium  microscopic  measures?  What do  the fluctuations
   in the vicinity of a stationary state look like?

 A large  amount of research  has been devoted to these questions  and
 to some  related ones. Although  the theory is far from being
 complete,  substantial progress has been made, particularly during
 the last twenty years. There are different ways to try to tackle the
 profound questions raised above. One line of research consists in
 exploring  structural properties of non-equilibrium systems: this
 endeavour   has led to celebrated results such  as  Fluctuation
 Theorems (Gallavotti and Cohen, 1995 \cite{Gallavotti}; 
 see also \cite{LeboSpohn}), 
 Non-equilibrium Work Relations (Jarzynski, 1997 \cite{Jarzynski})  or
 Macroscopic Fluctuation Theory (G. Jona-Lasinio and co-workers, see e.g.
 L. Bertini et  al. \cite{Bertini}).

 Another strategy to gain insight into non-equilibrium physics  is 
 to extract as much information as possible from analytical
 studies and from  exact solutions of some special  models.  The Ising
 model, that  has played a fundamental  role  in the theory
 of phase transitions, critical exponents and  renormalization group,
 is a landmark of  this style of  research.  In the field of
 non-equilibrium statistical mechanics, the Asymmetric Simple
 Exclusion Process (ASEP) is reaching  the  status of such  a
 paradigm. The ASEP consists of  particles   on a lattice, that
 hop  from a site  to its  immediate neighbours,   and
 satisfy the exclusion condition: a given location must  be occupied by
 at most one particle. Therefore, a jump  is allowed only if the
 target site is empty. Physically, the   exclusion constraint mimics short-range
 interactions amongst particles.  Besides, in order to drive this
 lattice gas out of equilibrium, non-vanishing  currents must  be
 established in the system. This can be achieved by various means: by
 starting from  non-uniform initial conditions, by coupling  the system
 to external reservoirs  that drive   currents \cite{Krug}  through the system
 (transport of particles, energy, heat) or  by introducing some
 intrinsic bias in the dynamics that favors motion in a privileged
 direction.  Then, each  particle is  an asymmetric random walker that
 drifts  steadily along the direction of an external driving force.
 Due to its simplicity, this model has appeared   in different
 contexts. It was  first proposed as a prototype to describe the
 dynamics of ribosomes along RNA \cite{McDonald1,McDonald2}. In the mathematical literature,
 Brownian processes  with hard-core interactions were defined  by
 Spitzer \cite{Spitzer}  who coined the name  exclusion process 
 (see also \cite{Harris,Liggett1,Liggett2}). The ASEP also
 describes  transport in low-dimensional  systems with
 strong  geometrical  constraints \cite{MartinRev1} such as  macromolecules transiting
 through  capillary vessels \cite{TomChou},  anisotropic conductors,  or quantum
 dots  where electrons hop to  vacant locations  and repel
 each other via  Coulomb interaction \cite{VonOppen}.  Very popular  modern
 applications of the exclusion process include molecular motors that
 transport proteins  along  filaments inside the cells  and, of
 course, ASEP and its variants are ubiquitous in discrete models of
 traffic flow \cite{MartinShock,Schad}. 
 More generally, the ASEP belongs to the class of driven diffusive systems
 defined by Katz, Lebowitz and Spohn in 1984 \cite{KLS}. 
  For a general discussion,
  we refer to  e.g., the book of H. Spohn \cite{Spohn} 
  the review    of B.~Schmittmann and R.  K. P. Zia \cite{Zia}
  and that of  G.~M.~Sch\"utz \cite{Schutz}.
  An early review of the  properties of the ASEP can
  be found in  Derrida, 1998 \cite{DerridaRep}.
  We emphasize that the ASEP is defined through dynamical rules: there
  is no energy associated with a microscopic configuration and thus
  there is no possibility of writing the stationary measure
  in the canonical form \eqref{Canonical}. More generally,
  the kinetic point of view seems to be a promising and
   fruitful approach to non-equilibrium systems: for such a presentation
 of statistical mechanics,  we highly  recommend to the reader the recent
 book by P. L. Krapivsky et al. \cite{PaulK}.

\begin{figure}[ht]
\begin{center}
   \includegraphics[height=1.75cm]{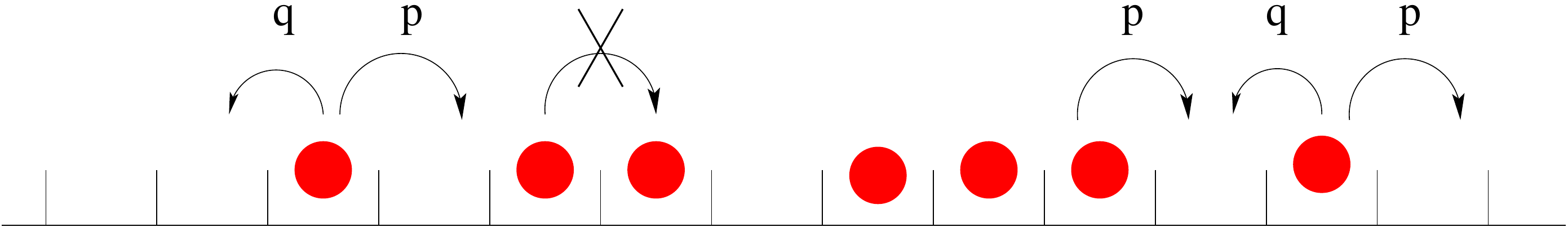}
  \caption{The Asymmetric Exclusion Process: 
  A  {\it paradigm}  for non-equilibrium
  Statistical Mechanics. The particles perform asymmetric random walks
 $(p \neq q)$  and interact through the exclusion constraint.}
\label{fig:ASEPgeneral}
 \end{center}
\end{figure}

  To summarize,
  the ASEP is a minimal model to study non-equilibrium behaviour
 (see Figure~\ref{fig:ASEPgeneral}). It is simple
 enough to allow  analytical studies, however  it contains the necessary
 ingredients for the emergence of a  non-trivial phenomenology \cite{Zia2}:
\begin{itemize}
 \item ASYMMETRIC: The  external driving  breaks detailed-balance and
 creates a stationary current in the system \cite{Zia3}.
  The model exhibits a non-equilibrium  stationary state. 

 \item EXCLUSION:  The hard core-interaction implies
 that there is  at most 1 particle per site. The ASEP is a genuine N-body problem.

\item PROCESS:   The dynamics is stochastic and  Markovian: there is  no
 underlying   Hamiltonian. 
\end{itemize}

The outline of this article is as follows. In section~\ref{sec:BA},
 we  study  spectral properties  of the ASEP
 on a ring by using the coordinate  Bethe Ansatz. We review the
 general technique and focus on the TASEP case, which,  in  our opinion,
 is one of the simplest models that allows to understand how the 
 Bethe Ansatz method works.  In section~\ref{sec:FluctCurrent},
 we explain how the fluctuations of the total current in the
 ASEP can be calculated using a functional formulation of the Bethe
 Ansatz; we review the  exact  results obtained and the different
 scaling regimes that appear in the limit of systems of large sizes.
 In section~\ref{sec:Multi}, we study a  generalization of the  ASEP
 in which different classes of particles interact through
 hierarchical dynamical rules. We show how  the steady state of such systems
 can be determined by using a Matrix Product Representation that involves
 tensor products of quadratic algebras.

 \section{Spectral Properties of the Markov Matrix}
  \label{sec:BA}

\subsection{The model}

We consider the exclusion process   on a periodic
one dimensional lattice with $L$ sites (sites $i$ and $L + i$ are
identical) and $N$ particles. 
 Because a  lattice site cannot be  occupied by more than  one particle, 
the  state of  a site  $i$ ($1 \le i \le L$)  can  be characterized
by  the Boolean number $\tau_i = 0, 1$ according as the site  $i$ is
 empty or occupied.

\begin{figure}[ht]
\begin{center}
   \includegraphics[height=4cm]{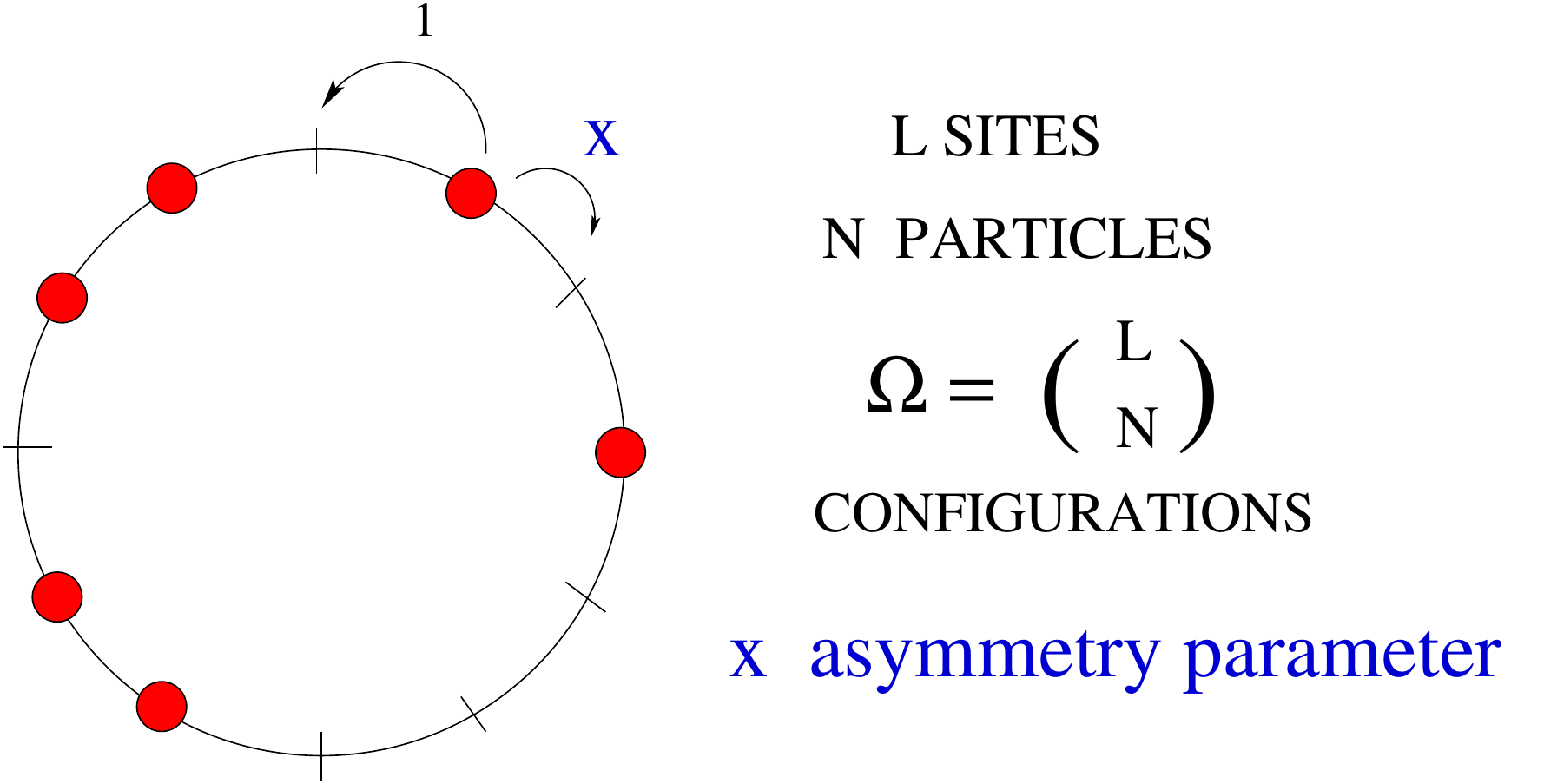}
  \caption{The Asymmetric Exclusion Process on a  Ring.}
\label{fig:ASEPring}
 \end{center}
\end{figure}

 The system evolves  in continuous time   according to 
the following stochastic rule: a particle on a site $i$ at time $t$ jumps, in
the interval between  $t$ and $t+dt$, with probability $ dt$ to the
neighbouring site $i+1$ if this site is empty ({\em exclusion rule})
 and with  probability $x\ dt$ to the
 site $i-1$ if this site is empty  (see Figure~\ref{fig:ASEPring}).  In 
  the totally asymmetric exclusion process (TASEP) 
 the jumps are totally biased in one direction ($x =0$). 
 On the other hand, the {\em symmetric} exclusion
  process (SEP) corresponds to the choice $ x = 1$.

The  number $N$ of particles is conserved by the dynamics. The total number of
configurations for $N$ particles on a ring with $L$ sites is given by
$\Omega = L! / [ N! (L-N)!]$.

A configuration $\mathcal{C}$  can be represented  by 
 the sequence $(\tau_1, \tau_2, \ldots, \tau_L).$ We call $P_t(\mathcal{C})$
 the probability of configuration $\mathcal{C}$ at time $t$.
  As the exclusion process is a continuous-time Markov process, the time
 evolution of $P_t(\mathcal{C})$ is determined by the master equation 
\begin{equation}
    \frac{d}{dt} P_t(\mathcal{C})  = \sum_{\mathcal{C}'}
      M(\mathcal{C},\mathcal{C}') P_t(\mathcal{C}')  \, . 
  \label{eq:Markov}
\end{equation}
 The   Markov  matrix  $M$  encodes  the dynamics of the exclusion process:  
 the element  $ M(\mathcal{C},\mathcal{C}')$
 is the  transition rate  from configuration $\mathcal{C}'$ to $\mathcal{C}$
 and  the diagonal term $M(\mathcal{C},\mathcal{C}) =
  -  \sum_{\mathcal{C}'}  M(\mathcal{C}',\mathcal{C})$ 
represents  the exit rate
 from configuration $\mathcal{C}$.

 A right eigenvector $\psi$ is associated with the eigenvalue $E$
of $M$ if
\begin{equation}
  M \psi = E\psi      \, . 
  \label{eq:mpsi=epsi}
\end{equation}

 The   matrix $M$ is a real non-symmetric matrix and, therefore, its
 eigenvalues (and eigenvectors) are either real numbers or complex
 conjugate pairs. The spectrum of $M$ contains the eigenvalue $E = 0$ and
 the associated right eigenvector is the stationary state. For the ASEP
 on a ring the steady state    is uniform and  all configurations
 have the same probability $1/\Omega$ \cite{DerridaRep}.

  Because the dynamics is ergodic ({\it i.e.}, $M$ is 
irreducible and aperiodic), the Perron-Frobenius theorem
 (see, for example, Gantmacher 1959 \cite{Gantmacher})
implies that 
 0 is a non-degenerate eigenvalue and that all other 
eigenvalues $E$  have a strictly negative real
part; the relaxation time of the corresponding eigenmode is $\tau =
-1/\mathrm{Re}(E)$. 
 The imaginary part of $E$ gives rise to an oscillatory  behaviour.

\begin{figure}[ht]
     \begin{center}
    \includegraphics[width=12.0cm]{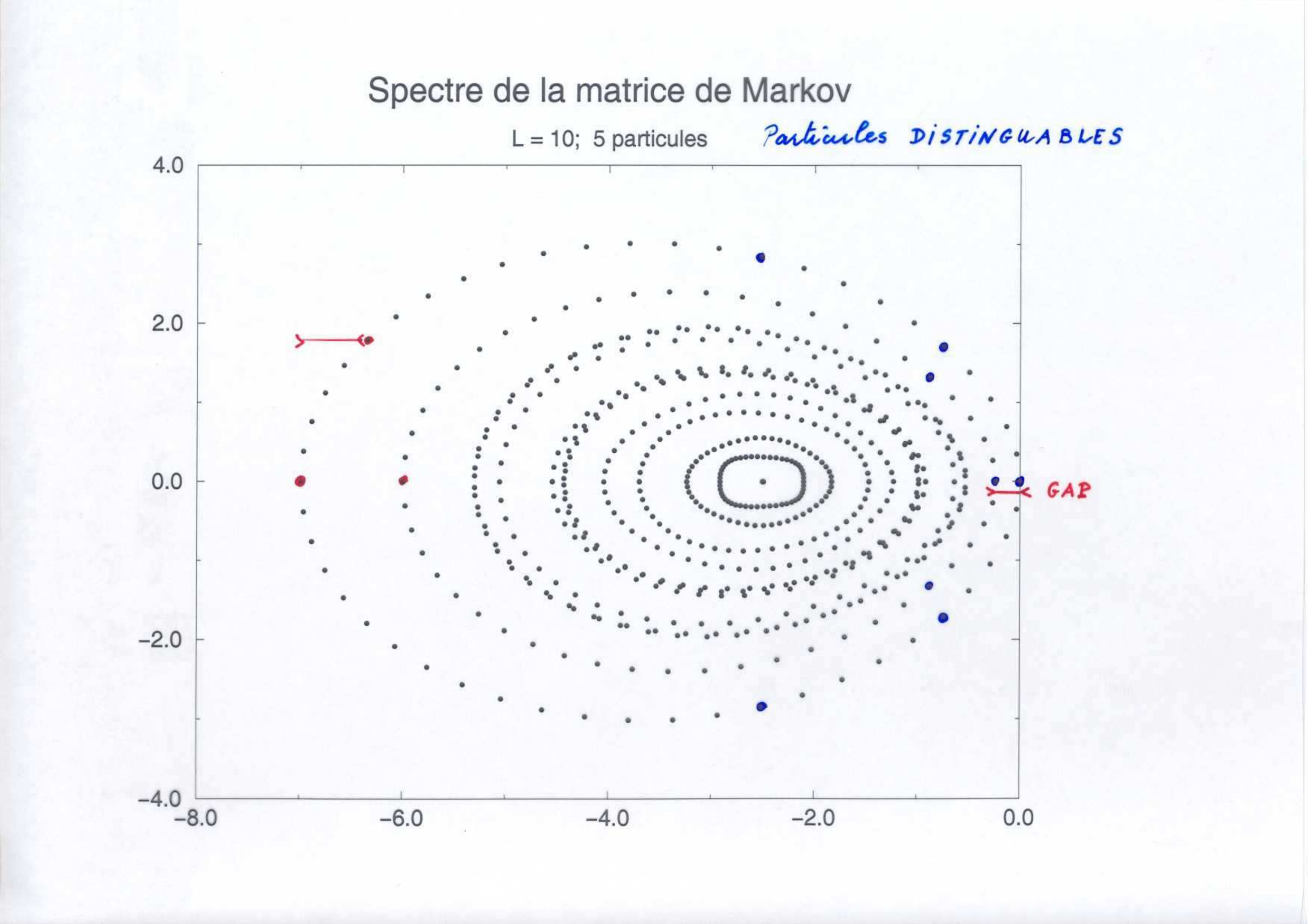} 
\label{fig:spectre}
 \caption{Example of a spectrum of a Markov matrix for the  TASEP with $N=5$
 distinguishable particles and a ring of $L=10$ sites.
 The thick dots indicate some of the  eigenvalues that 
 correspond  to the   indistinguishable case.}
     \end{center}
    \end{figure}

   In  Figure~4,    we display the example of 
  a spectrum of a Markov matrix for the  TASEP with $N=5$
 distinguishable particles and a ring of $L=10$ sites.  In this case,
  the dimension  of the phase space is  1260. However,  when the
    particles  are indistinguishable, the  dimension of  the   Markov matrix
  is reduced  by a factor of  $N=5$  and  is given by 252. The spectrum
  of the  indistinguishable case is included in the larger spectrum of the 
  distinguishable problem.

 \subsection{The Bethe Ansatz}

 Another way to characterize  a configuration is to specify 
  the positions of the $N$ particles
on the ring, $(\xi_1, \xi_2, \dots, \xi_N)$ 
 with $1 \le \xi_1 < \xi_2 < \dots < \xi_N \le L$.
 In  this representation, the eigenvalue equation
 (\ref{eq:mpsi=epsi})   becomes
\begin{eqnarray}
 &&  E \psi(\xi_1,\dots, \xi_N) =   \nonumber \\   &&
  \sum_i  \left[ 
    \psi(\xi_1, \dots, \xi_{i-1},\ \xi_i-1,\ \xi_{i+1}, \dots, \xi_N) 
 - \psi(\xi_1,\dots, \xi_N) \right]
    \nonumber \\      
+  &&  \sum_j  x 
 \left[ \psi(\xi_1, \dots, \xi_{j-1},\ \xi_j+1,\ \xi_{j+1}, \dots, \xi_N)
              - \psi(\xi_1,\dots, \xi_N) \right] \, , 
\label{eq:eigeneq}
\end{eqnarray}
where the sums run  over the indexes $i$ such that $ \xi_{i-1} < \xi_i-1$
 and   over the indexes $j$ such that $  \xi_j + 1 < \xi_{j+1} $. In other words, 
 the sums are restricted to   jumps  that  respect the exclusion condition \cite{DerridaRep}.

Since the works of Dhar 1987 \cite{Dhar}, and  Gwa and Spohn  1992 \cite{Gwa},
 it is known that the   Bethe  Ansatz  can be applied to the ASEP.
The idea of the  {\em  Bethe  Ansatz} (H. Bethe 1931 \cite{Bethe})  consists
 in writing the eigenvectors   $\psi$  of
the Markov matrix as linear combinations of plane waves:
\begin{equation}
  \psi(\xi_1,\dots,\xi_N) = \sum_{\sigma \in \Sigma_N} A_{\sigma}  \,  
  z_{\sigma(1)}^{\xi_1} \,  z_{\sigma(2)}^{\xi_2} \dots z_{\sigma(N)}^{\xi_N}
  \label{eq:ba} \, , 
\end{equation}
where $\Sigma_N$ is the  group of the $N!$ permutations of $N$ indexes. The
coefficients $\{A_{\sigma}\}$ and the wave-numbers $\{z_1, \dots, z_N\}$
are complex numbers to be determined.
 We  observe that in the case where all  particles are
 far from each other (i.e. no particles are  located on 
  adjacent sites so that 
   $ \xi_{k-1} < \xi_k-1$ for all $k=1,\ldots N$),
 each monomial that appears  on the right hand
 side of the expression of the Bethe wave function is a solution
 of the 
 eigen-equation~\eqref{eq:eigeneq}, with an  eigenvalue $E$ given by
\begin{equation}
 E(z_1, z_2 \ldots  z_N ) =    \sum_{i =1}^N 
   \frac{1}{ z_i}  + 
        x  \sum_{i =1}^N   z_i \, - N (1+x)\, . 
  \label{eq:eigenval} 
\end{equation}
 However, in order for the trial wave function to be a genuine  eigenfunction,
 equation~\eqref{eq:eigeneq} must be fulfilled  for all configurations:
 therefore, 
 one has to insure that when two or more  particles are adjacent
  equation~\eqref{eq:eigeneq} is still satisfied.  The case where exactly 
 two particles are adjacent  and all other  particles are well separated
 from one another is enough to fix the value of all the 
 $A_{\sigma}$'s (up to an overall multiplicative constant). In other words, 
  the form of the Bethe  wave-function~\eqref{eq:ba} is fully 
  determined by the
  `two-body collisions'. However, more  particles can form larger clusters:
 this corresponds to $k$-body collisions with $2 < k \le N$: 
  such collisions  impose further  constraints that the Bethe  wave-function
  has no reasons, a priori, to satisfy. Fortunately,   in the present problem,
  the constraints imposed by the  $k$-body collisions can be written as  linear
  combinations of the  two-body constraints and  are therefore
   automatically satisfied by the Bethe  wave-function. 
 In the Bethe Ansatz jargon,  this fact  is formulated by saying
 that  the $3$-body collisions {\it factorize} into
 $2$-body collisions etc... This  remarkable property, that 
 can be verified explicitly in the case of  equation~\eqref{eq:eigeneq}, lies
 at the heart of the   {\it integrability} of the exclusion process, i.e., 
 it implies that the ASEP can be  solved  by Bethe Ansatz. 
 In fact,  the ASEP  can be mapped exactly into well-studied
  systems such as quantum spin chains \cite{Alcaraz,Albertini},
  vertex models \cite{Baxter,Kandel,VanBeij2} or solid-on-solid models \cite{Rajesh}
  which  are well-known to be integrable.

Because we are studying the system on a homogeneous  ring, the 
 function $\psi$ must also satisfy the following periodic boundary conditions
\begin{equation}
  \psi(\xi_1, \xi_2, \dots, \xi_n) = \psi(\xi_2, \dots, \xi_n, \xi_1 + L) \, .
  \label{eq:cycl}
\end{equation}
 These   periodic conditions 
 quantify the eigenvalues by imposing  a set of equations 
 satisfied by  the $z_i$'s,
 the  {\it  Bethe Equations}:
\begin{equation}
  z_i^L  = (-1)^{N-1} \prod_{j =1}^N
   \frac{x   z_i z_j - (1+x) z_i +  1}
  {x   z_i z_j -(1+x) z_j + 1  } \, . 
  \label{eq:BetheEq} 
\end{equation}
 This  set of non-linear algebraic
 equations  must be satisfied by the fugacities. Therefore, in order to obtain  the spectrum
 of the ASEP   Markov matrix  one has  first  to find all
 $N$-tuplets $(z_1,z_2,\ldots,z_N)$  that solve the  Bethe
  equations~\eqref{eq:BetheEq}. Then,  given a solution,  one can calculate
 the corresponding eigenvalue~\eqref{eq:eigenval} 
   and  eigenfunction~\eqref{eq:ba}. Of course, in practice
 this program  can be carried out only in  some special situations 
 (for example in  a limited portion of the  spectrum) and usually one has to take 
  the `thermodynamic'
 limit $L,N \to \infty$ with a  fixed value
 of the density $\rho = N/L$.  Besides, the  completeness  issue  (i.e. whether 
  the Bethe Ansatz does  provide the  full spectrum)
 is a difficult problem  in  algebraic geometry \cite{Baxtervertex,Langlands}. 

\subsection{Bethe Equations for the TASEP} 

   The totally asymmetric exclusion process
  (TASEP), that corresponds to  $x = 0$,   provides a particularly
  instructive  illustration  of   the Bethe Ansatz. The Bethe
  equations take a  simpler form on which   analytical
  calculations can be carried out even for finite values of  $L$ and $N$.  
  The TASEP  is one of the simplest non-trivial models
  that allows to  understand  how the Bethe Ansatz technique works.
  In this  subsection, we present a complete and
  self-contained derivation  of the Bethe Ansatz for the TASEP.

  First,  we show that the
  Bethe wave function $\psi$ of the TASEP can be written as a
  determinant \cite{ogkm2}. The form of this determinant can be guessed
  heuristically by working out explicitly  examples for  small systems. Here,
  we   reverse  the logic and   {\it define}  $\psi$ as
\begin{equation}
    \psi(\xi_1,\dots,\xi_N) = \det(R)  \, , 
   \label{eq:psidet}
\end{equation}
where $R$ is a $N \times N$ matrix with elements
\begin{equation}
   R(i,j) = \frac{z_{i}^{\xi_j}}{(1-z_i)^j}
  \ \ \mbox{for } 1 \le i,j \le N  \, ,
   \label{eq:r}
\end{equation}
$(z_1, \dots, z_N)$ being $N$  complex numbers. By expanding the
 determinant, one recovers the generic form~\eqref{eq:ba} for
 the   Bethe wave function $\psi$. We now show that  $\psi$, thus  defined,
 is a  solution of  the eigenvalue equation~\eqref{eq:eigeneq} with $x =0$.
 First, we  need to prove that  $\psi$
 satisfies two identities which are valid for any values of
$z_i$ and of $\xi_i$.    The first identity is
\begin{equation}
  E \psi(\xi_1,\dots, \xi_N) = \sum_{k=1}^N
   [\psi(\xi_1,\dots,\xi_k-1,\dots, \xi_N) - \psi(\xi_1,\dots,\xi_k,\dots, \xi_N)] \, , 
  \label{eq:evp}
\end{equation}
  where $E$ is given by  $E = -N + \sum_{i=1}^N 1/z_i\,.$ We emphasize  that
 this equality is true 
for any $(\xi_1, \dots, \xi_N)$ and $(z_1, \dots, z_N)$
 without imposing the ordering
   $1 \le \xi_1 < \xi_2 < \dots < \xi_N \le L$. Besides, 
  $k$ runs from 1 to $N$ without  any restriction, i.e. we do {\it  not} impose
   that $ \xi_{k-1} < \xi_k-1$. 
Equation~(\ref{eq:evp}) is proved  by writing
\begin{eqnarray}
  \psi(\xi_1,\dots,\xi_k-1,\dots, \xi_N) - \psi(\xi_1,\dots,\xi_k,\dots, \xi_N) 
   =  \nonumber \\
  \det \left( R(i,1) , \dots ,\left( \frac{1}{z_i} -1 \right) R(i,k) ,
           \dots , R(i,N)
        \ \right) \, .
\end{eqnarray}
This determinant is similar to $ \det(R)$ except for the $k$-th
column.  Expanding this determinant over all permutations of $\{ 1,
\dots, N \}$ and performing the sum over $k=1,\dots,N$ leads to the desired
equation.
The second identity takes care of the two  particles collision case: it is 
  valid for any $(z_1, \dots, z_N)$ and any $(\xi_1,
\dots, \xi_N)$ with $\xi_{k-1} = \xi_k$, and is given by 
\begin{equation}
  \psi(\xi_1,\dots, \xi_k, \xi_k, \dots, \xi_n) - 
  \psi(\xi_1,\dots, \xi_k, \xi_k+1, \dots, \xi_n)  = 0 \, .  
  \label{eq:kk+1}
\end{equation}
This  identity  is proved as follows: we rewrite the 
 left hand side of equation~(\ref{eq:kk+1})   as the determinant 
$\det(\tilde{R})$ where $\tilde{R}$ is a matrix that is identical to
$R$ except for its $k$-th column that is given by
\begin{equation}
  \tilde{R}(i,k) = \frac{z_i^{\xi_k} - z_i^{\xi_k+1}}{(1-z_i)^{k}} = 
                \frac{z_i^{\xi_k}}{(1-z_i)^{k-1}} = R(i,k-1) =  \tilde{R}(i,k-1) \,.
\end{equation}
 We remark that the 
  $(k-1)$-th and the $k$-th columns of $\tilde{R}$ are equal and,
therefore, $\det(\tilde{R}) = 0$, proving equation~(\ref{eq:kk+1}).

        To conclude, we note that the  eigenvalue equation~\eqref{eq:eigeneq}
 is similar  to  equation~(\ref{eq:evp}) except that in~\eqref{eq:eigeneq} 
 the sum is restricted to the
allowed jumps of particles, i.e., to the values of  $k$ such that
$\xi_{k-1} +1 < \xi_{k}$.  However, in equation~(\ref{eq:evp}), the terms
with $\xi_{k-1}+ 1 = \xi_{k}$ vanish thanks to equation~(\ref{eq:kk+1}).
Hence,  equation~(\ref{eq:evp}) is  in fact exactly the same as  the eigenvalue equation
when  the eigenvector has the form assumed in equations~(\ref{eq:psidet},
\ref{eq:r}).

         Finally, because of the  periodic boundary conditions~\eqref{eq:cycl}, the 
$z_i$'s  are quantified by  the  Bethe equations, that we now rederive.
 Denoting by $i$ and $j$ the generic line
and column  of the matrix $R$, we can write 
\begin{equation}
  \psi(\xi_2, \dots, \xi_N, \xi_1 + L) = \det \left( 
       \frac{z_{i}^{\xi_{2}}}{1-z_i}, 
     \dots,  \frac{z_{i}^{\xi_{j+1}}}{(1-z_i)^j} , \dots,
             \frac{z_i^{\xi_1+L}}{(1-z_i)^N}  \right) \, .
  \label{eq:pbc}
\end{equation}
By cyclic permutation of the columns, we obtain
\begin{eqnarray}  
  & &  \psi(\xi_2, \dots, \xi_N, \xi_1 + L)     \nonumber  \\
    &=&   (-1)^{N-1} 
    \det \left( \frac{z_i^{\xi_1+L}}{(1-z_i)^N},  \frac{z_{i}^{\xi_{2}}}{1-z_i}, 
     \dots,  \frac{z_{i}^{\xi_j}}{(1-z_i)^{j-1}} , \dots   \right)  \nonumber \\
     &=&   (-1)^{N-1} 
    \det \left( \frac{z_i^L}{(1-z_i)^{N-1}} \, R(i,1),
     \dots,  (1-z_i)\, R(i,j) , \dots \right)    \nonumber  \\
    &=&   (-1)^{N-1}  \prod_{k=1}^N (1-z_k)  \,\,\, \
    \det \left( \frac{z_i^L}{(1-z_i)^N} \, R(i,1),
     \dots,  R(i,j) , \dots   \right)     \, .  
  \end{eqnarray}
  The last expression  will be  equal to
  $\psi(\xi_1, \xi_2, \dots, \xi_n) = \det(R) \, $ if
  $z_1,\dots, z_N$ are solutions of the  TASEP  Bethe equations:
\begin{equation}
  (z_i-1)^N z_i^{-L} = - \prod_{k=1}^N(1-z_k) \ \ \
   \hbox{for}    \ \ \  i=1,\dots,N  \, .
 \label{eq:beTASEP}
\end{equation} 
These equations can  of course be  obtained by substituting $x=0$ in the
 general Bethe equations~\eqref{eq:BetheEq} for ASEP. We note
 that determinantal representations of the eigenvectors
 and of the exact probability distribution at finite time play
 an important role in the study of the TASEP since the seminal work of G. M.
 Sch\"utz \cite{SchutzBA}, that has been generalized further by 
   V.~B.~Priezzhev   \cite{Priezzhev} and N.~M.~Bogoliubov   \cite{Bogoliubov}
  (see e.g. the review
 of  V.~B.~Priezzhev 2005  \cite{PriezzhevInde}).

\subsection{Analysis of the TASEP  Bethe Equations} 
\label{sec:procedure}

\begin{figure}[ht]
\begin{center}
   \includegraphics[height=5.4cm]{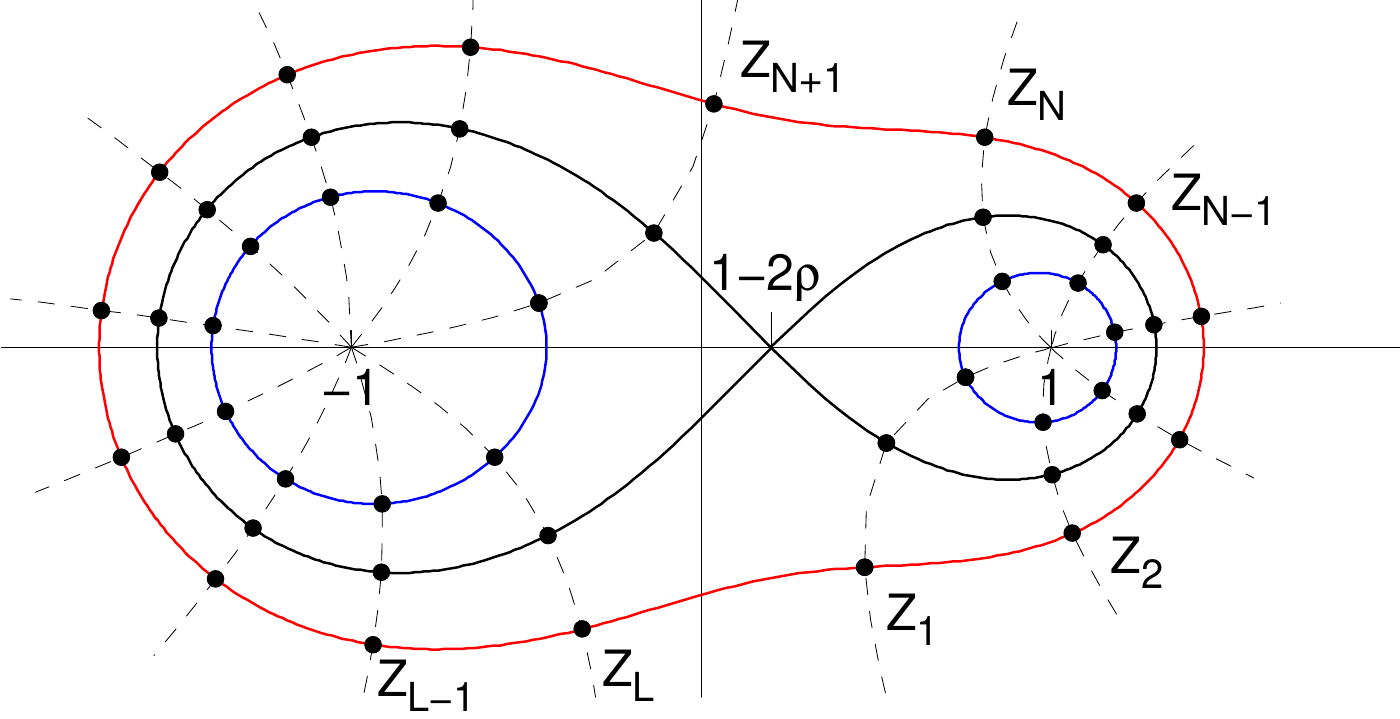}
  \caption{The loci of the roots of the Bethe Equations for the TASEP
 are given by Cassini Ovals.}
\label{fig:cassini}
 \end{center}
\end{figure}

  The Bethe equations for the TASEP exhibit a remarkable `decoupling'
 property that allows to perform analytical calculations even for finite
 values of $L$ and $N$. Using the auxiliary variables 
 $Z_i = 2/z_i -1$, the Bethe equations~\eqref{eq:beTASEP}  become 
 \begin{equation}
  (1-Z_i)^N \ (1+Z_i)^{L-N}  
  =  - 2^L \prod_{j=1}^N \frac{Z_j - 1}{Z_j + 1}  \ \ \
  \hbox{for}    \ \ \ i=1,\dots,N     \, .
  \label{eq:bez}   
\end{equation} 
 We note that the right-hand side of these equations is independent
 of the index $i$: this property is true only for the TASEP
 where  the  Bethe equations decouple and can be  reduced
 to a   polynomial  in one-variable plus  a self-consistency
  equation as shall be explained below. 
 The corresponding eigenvalue $E$ of  the Markov matrix  $M$ is given by 
 \begin{equation}
 2  E =     -N + \sum_{j=1}^N Z_j  \, .
 \label{eq:eigenvalTASEP}
 \end{equation}
The   solutions of the   Bethe  equations~(\ref{eq:bez})
 are the roots  of the  polynomial equation of  degree $L$ 
\begin{equation}
   (1-Z)^N (1+Z)^{L-N} = Y \, , 
  \label{eq:zzy}
\end{equation}
 where  $Y$ is   determined self-consistently
 by the r.h.s. of  equation~(\ref{eq:bez}). 
  Given   an arbitrary 
 value of the  complex number $Y$, the roots 
 of the  polynomial~(\ref{eq:zzy}) display a simple 
 geometrical layout. If we write   $Y = r^L \, e^{i \phi}$, 
$r$ being a positive real number, we observe that 
 equation~(\ref{eq:zzy}) implies that 
\begin{equation}
   |Z-1|^{\rho} |Z+1|^{1-\rho} = r
 \label{eq:Cassini}
\end{equation}
where $\rho = N/L$ is the particle density in  the system. 
The  curves defined in the complex plane for $ 0 < r < \infty$
 are known as  {\it Cassini ovals} (see Figure~\ref{fig:cassini}).
  As can be seen in 
 Fig.~\ref{fig:cassini}, the topology of the Cassini ovals  depends on
the value of $r$ with a critical value:
\begin{equation}
   r_c = 2 \rho^\rho(1-\rho)^{1-\rho} \, .
   \label{eq:rc}
\end{equation}
\hfill\break
  For $r < r_c$, the curve consists of two disjoint ovals with $N$
  solutions on the oval surrounding $+1$ and $L-N$ solutions on the
  oval surrounding $-1$.
  \hfill\break
 For  $r = r_c$, the curve is a deformed Bernoulli lemniscate
  with a double point at $Z_c = 1 - 2 \rho$.
  \hfill\break
  For $r > r_c$, the curve is a single loop with $L$ solutions. \hfill\break
  Note that the
  Cassini ovals are symmetrical only if $\rho = 1/2$; in this case 
 $r_c = 1.$

\hfill\break

  This geometric layout leads to
 the following procedure for solving the Bethe Equations for the TASEP
 (for a review see e.g.  \cite{ogkmrev}):
\begin{itemize}
\item[$\bullet$] {\it SOLVE},  for any given value of $Y$,  the equation 
        $ \,\,\, ( 1 -z_i)^N   ( 1 +z_i)^{L-N}  =  Y \, . $
   The roots are located on    {Cassini Ovals} 
\item[$\bullet$]   {\it CHOOSE   $N$ roots  }
  $z_{c(1)}, \ldots z_{c(N)}$ amongst  the $L$ available roots,
 with  a  {\it choice set} $c : \{ c(1),\ldots , c(N)\} \subset
 \{ 1,\ldots ,L\}  \, . $
\item[$\bullet$] {\it SOLVE }  the  {self-consistent}
 equation ${\bf   A_c(Y) = Y }$  where
          $$A_c(Y) =  -2^L \prod_{j=1}^N  
 \frac{ z_{c(j)} -1}{ z_{c(j)} + 1}  \, .$$
\item[$\bullet$] {\it DEDUCE  }   from the value of $Y$,   the 
  $z_{c(j)}$'s and the energy corresponding to the choice 
  set  $c$~:
       $$ 2E_c(Y) = -N + \sum_{j=1}^N z_{c(j)} .$$
\end{itemize}
   Eigenvalues of the Markov matrix are thus related to 
  the   choice sets of roots on the Cassini ovals; 
  there is a strong evidence  \cite{ogkmrev}  that  choice sets
 are in one-to-one correspondence with the eigenvalues
 of the  Markov matrix. A simple argument is  that the total number
 of  choice sets is given by $\Omega = L! / [ N! (L-N)!]$ which is precisely the 
 size of the  Markov matrix.

 The choice function $c_0(j) = j$ that selects the $N$ fugacities $Z_i$ with
 the largest real parts (see Figure~\ref{fig:cassini})  provides the ground state of the Markov
 matrix. The spectral gap, given by the 
  first excited eigenvalue,  corresponds to the choice 
  $c_1(j) = j$ for $j = 1, \ldots, N-1$ and $c_1(N) = N+1$ \cite{Gwa}. 
  For this choice set, the calculation  can be carried out explicitly and one can show
  that, in the large $L$ limit, the  first excited state  is  given 
 by a  solution of a  transcendental equation.
  For a density $\rho$, one obtains
\begin{eqnarray}
    E_1  = 
 &&  { -2 \sqrt{\rho( 1 - \rho)} \frac{6.509189337\ldots}{L^{3/2}} }
      \pm  {\frac{ 2 i \pi (2 \rho -1)}{L}  \, .}  \nonumber \\
          &&   \,\,\,\,\,\,  \,\,\,\,\,\,  \,\,\,\,  {\rm {(RELAXATION)}}  
         \,\,\,\,  \,\,\,\, \,\,\,\, \,\,\,\, \,\,\,\, 
 {\rm  {(OSCILLATIONS)} }      \nonumber 
\end{eqnarray}
   The first excited state  consists
     of a  pair of conjugate  complex numbers when 
   $\rho$  is different from 1/2. The real part of $E_1$ describes the relaxation
    towards the stationary state: we find that   the largest relaxation time  
    scales as  $T \sim L^z$   with the dynamical exponent 
   $z=3/2$  \cite{VanBSpohn,Dhar,Gwa,Kim1,ogkm1}.   This value
  agrees with the dynamical exponent of the   one-dimensional 
 Kardar-Parisi-Zhang equation   that belongs to the same universality
 class  as ASEP (see the review of Halpin-Healy and Zhang 1995 \cite{HHZ}). The imaginary part of 
  $E_1$  represents the  relaxation oscillations and scales as $L^{-1}$;
 these  oscillations correspond to a kinematic wave that propagates with
  the group velocity $2 \rho -1$:  such    traveling waves  can be 
  probed by dynamical  correlations \cite{Majumdar,Barma}.  
    
   The same procedure also allows
  to classify the higher excitations in the spectrum \cite{deGier2}.
  For the  partially asymmetric case
 ($x \neq 0$),  the Bethe equations do not decouple and analytical
 results are much harder to obtain. A systematic procedure for calculating the
 finite size corrections of the upper region of
  the ASEP spectrum was developed by Doochul Kim \cite{Kim1,Kim2}.

  We  emphasize that the results presented here are valid for the ASEP
  on a finite periodic lattice. 
  Problems with  open  boundary conditions   \cite{deGier1,deGier2,deGier3}
  or on  infinite lattices \cite{Sasamoto}  require  different approaches.
  In particular, the  exclusion process
  on the infinite line  has fascinating  connections
  with  random matrix theory and with 
  the   Robinson-Schensted-Knuth correspondence. 
 More precisely, consider the {\it totally} asymmetric case  and 
 suppose that the initial configuration of the TASEP 
  is such that all sites $i \le 0$ are occupied and  all 
  sites with   $i >  0$ are empty.  Then,  the probability that a particle
  initially at $-m$ moves at  least $n$ steps to the right in time $t$
 equals the probability distribution of the largest eigenvalue  in a unitary
 Laguerre random matrix ensemble. This crucial fact,  first understood
 by Johannson in  2000 \cite{Johansson}, allows to relate  the statistical
 properties of the {\it totally} asymmetric exclusion process 
 to the  classical  Tracy-Widom laws  \cite{TracyWidom}
 for   the largest eigenvalue in  ensembles  of random
 matrices.  The study  of such   relations  has become 
   a subfield per se and  has  stimulated many   works
(see e.g.,  \cite{SpohnInde,SpohnCours,FerrariPatrick}). 
 In particular, the  determinantal representation of the transition
 probabilities of   the TASEP allows to retrieve  Johannson's result 
 in a very  appealing manner \cite{Rakos}.  More recently,
 in a  series of articles  \cite{Tracy1,Tracy2,Tracy3,Tracy4,Tracy5}, 
 C. A. Tracy and H. Widom have  found  some integral formulas
 for the  probability distribution of an individual
 particle  starting from step initial conditions
 in  the asymmetric exclusion process with general parameter values 
 (i.e. allowing  forward and backward jumps). 
 These expressions can be rewritten as 
  a single integral involving a Fredholm determinant, that is amenable
  to   asymptotic analysis. In particular,   a limit theorem is proven 
  for the total current distribution. These breakthroughs  extend
 the results of Johansson on TASEP to ASEP: this is a crucial progress 
 because  the weakly asymmetric process leads to a well-defined
 continuous limit.  In particular,  a  very
 important outgrow of these studies are the recent
 papers of  H. Spohn, T. Sasamoto
 and S. Prolhac, in which the height fluctuations of the Kardar-Parisi-Zhang
 equation and $n$-point correlation functions are exactly derived, solving
 a problem that remained open for 25 years (see the articles
  of   H. Spohn and T. Sasamoto, and of P. L.  Ferrari in the present
 special issue). 
  For an overview 
 of all these fascinating problems,  we refer the reader  to  the 
  recent review article by  Kriecherbauer and Krug 2010 
  \cite{Kriecherbauer}.

  Finally, we note that the Bethe Ansatz can be 
  used  in models that are closely related to the ASEP, such
 as the zero-range process \cite{Povolotsky}, the
  raise and peel model \cite{AlcarazRitt},  
 vicious walkers \cite{Dorlas}  or the
 Bernoulli matching model of sequence  alignment \cite{PriezSchutz}.

 \section{Fluctuations of the total  current}
 \label{sec:FluctCurrent}

   In this section, we explain how the  statistics  of the  total current 
  \cite{FerrariFontes,VanBeij3} 
  in the ASEP on a  ring can be determined by the  Bethe Ansatz. 
 In particular, we show  that  the fluctuations of the  current 
 can display a  non-Gaussian  behaviour
   in contrast with the  equilibrium case.

 \subsection{Current statistics and Bethe Ansatz}

 We  call $Y_t$ the total distance
 covered by all the particles between time 0 and time $t$
 and  define $P_t(\mathcal{C}, Y)$ the joint probability
 of being   in the configuration $\mathcal{C}$ at time $t$ and
 having $Y_t = Y$. The evolution equation of  $P_t(\mathcal{C}, Y)$ is:
\begin{equation}
    \frac{d}{dt} P_t(\mathcal{C}, Y)  = \sum_{\mathcal{C}'} \Big(
   M_0(\mathcal{C},\mathcal{C}') P_t(\mathcal{C}', Y)  
+   M_1(\mathcal{C},\mathcal{C}')  P_t(\mathcal{C}', Y -1)  
+   M_{-1} (\mathcal{C},\mathcal{C}')  P_t(\mathcal{C}', Y +1)  
        \Big)   \, . 
 \label{eq:Markov2}
\end{equation}
 Using  the generating function  $F_t(\mathcal{C})$
 \begin{equation}
  F_t(\mathcal{C}) =  \sum_{ Y =0}^\infty 
  {\rm e}^{\gamma Y} P_t(\mathcal{C}, Y) \, ,
 \label{eq:defF}
\end{equation}
  the evolution equation becomes
\begin{equation}
    \frac{d}{dt} F_t(\mathcal{C})  = \sum_{\mathcal{C}'} \Big(
   M_0(\mathcal{C},\mathcal{C}') 
+ {\rm e}^\gamma  M_1(\mathcal{C},\mathcal{C}')  
+   {\rm e}^{-\gamma}  M_{-1} (\mathcal{C},\mathcal{C}')   
        \Big)  F_t(\mathcal{C}') =  \sum_{\mathcal{C}'} 
  M(\gamma)(\mathcal{C},\mathcal{C}')   F_t(\mathcal{C}')
   \, . 
 \label{eq:Markov3}
\end{equation}
This equation is   similar to the original 
 Markov  equation~(\ref{eq:Markov}) for the probability
 distribution  $ \psi_t(\mathcal{C})$ but now  the original
 Markov matrix $M$ is deformed  by a  jump-counting   {\it  fugacity } ${\gamma}$ into 
  $M(\gamma)$  (which  is  not a Markov matrix in general),  given by
 \begin{equation}
  M(\gamma) =   M_0 +  {\rm e}^\gamma  M_1 +  {\rm e}^{-\gamma}  M_{-1} \, . 
 \label{eq:defMgamma}
\end{equation}
 In the long time limit, $ t \to \infty$, 
  the behaviour of   $F_t(\mathcal{C})$
 is dominated by the largest eigenvalue  $E(\gamma)$  and one can write 
  \begin{equation}
  {  \left\langle  {\rm e}^{\gamma Y_t}   \right\rangle \simeq 
    {\rm e}^{ E(\gamma) t} }  \, . 
 \label{eq:limF}
\end{equation}
 Thus,  in the long time limit, the   function $E(\gamma)$  is the generating function of 
 the  cumulants  of the total current $Y_t$.  But   $E(\gamma)$  is also the
 dominant  eigenvalue of the   matrix   $M(\gamma)$. Therefore, 
  the current statistics has been traded into  an eigenvalue problem. Fortunately,  the deformed matrix 
  $M(\gamma)$  can still be diagonalized by the Bethe Ansatz.
 In fact, a  small modification of the calculations described in Section~\ref{sec:BA}
 leads  to the following  Bethe Ansatz equations 
\begin{equation}
  z_i^L  = (-1)^{N-1} \prod_{j =1}^N
   \frac{x {\rm e}^{-\gamma}  z_i z_j - (1+x) z_i + {\rm e}^{\gamma}  }
  {x{\rm e}^{-\gamma}  z_i z_j -(1+x) z_j +{\rm e}^{\gamma} }  \, .
\label{eq:BAgamma}
\end{equation}
 The  eigenvalues of $M(\gamma)$  are   given by 
\begin{equation}
 E(\gamma; z_1, z_2 \ldots  z_N ) =  {\rm e}^{\gamma}  \sum_{i =1}^N 
   \frac{1}{ z_i}  + 
        x {\rm e}^{-\gamma}  \sum_{i =1}^N   z_i \, - N (1+x)\, .
\label{eq:Egamma}
\end{equation}
 The cumulant generating function corresponds to the largest eigenvalue.

\hfill\break
  {\it Remark:}  One can  define 
 $ { G(j) }$,  the  {\it  large-deviation function}  of the current,
   as follows 
\begin{equation} 
 Prob\left(\frac{Y_{t}}{t}=j\right) {\sim}e^{-tG(j)} \, .\label{def:LDF}
\end{equation}
 It  can be shown using~\eqref{eq:limF} that  $G(j)$ and $E(\gamma)$
  are  the {  Legendre transforms } of each other.
 Large deviation functions play an increasingly important role in
 non-equilibrium statistical physics (see  H. Touchette 2009 \cite{Touchette}),
 in particular in relation to the Fluctuation Theorem \cite{LeboSpohn}. Thus, 
 determining exact expression for these  large deviation functions for interacting
 particle processes, either analytically or numerically is a major challenge in the field 
 (we refer the reader to the review of B. Derrida, 2007  \cite{DerrReview}). Besides,
 higher cumulants of the current
 and  large deviations are  also of experimental interest in relation to  counting
  statistics in quantum systems \cite{Flindt}.

\vskip 0.5cm

\subsection{The TASEP Case}

  For the  TASEP, the  Bethe equations~\eqref{eq:BAgamma}  decouple and can be studied
 by  using the  procedure outlined in  Section~\ref{sec:procedure}. 
  The  $x=0$ case was completely  solved by   B. Derrida and J. L. Lebowitz in  1998 \cite{DLeb}. 
 These authors  calculated   $E(\gamma)$ 
 by Bethe Ansatz to all orders  in $\gamma$. More precisely,  they obtained 
 the following  representation
 of the function  $E(\gamma)$  in terms  of an auxiliary  parameter $B$: 
\begin{eqnarray}
      E(\gamma)  &=&  -N   \sum_{k=1}^\infty 
             \left(\begin{array}{c} kL-1\\kN \end{array} \right)
  \frac{B^k}{kL-1}    \, ,  \label{eq:EofY}\\
           \gamma  &=&  - \sum_{k=1}^\infty   
  \left(\begin{array}{c} kL\\kN \end{array} \right)  \frac{B^k}{kL} 
        \label{eq:gamY} \, . 
\end{eqnarray}
These expressions allow to calculate the cumulants of $Y_t$, for example the mean-current $J$
 and the diffusion constant $D$:
\begin{eqnarray}
J =  \lim_{t \to \infty} \frac{ \langle Y_t \rangle}{t} &=&
 \frac{ {\rm d}  E(\gamma)}{ {\rm d} \gamma} 
 \Big|_{\gamma =0} =  \frac{ N(L-N)}{L-1}\, ,    \\ 
 \,\,\, D = \lim_{t \to \infty} 
  \frac{ \langle Y_t^2 \rangle - \langle Y_t \rangle^2 }{t}  &=&
  \frac{ {\rm d}^2  E(\gamma)}{ {\rm d} \gamma^2} 
 \Big|_{\gamma =0} =
 \frac{N^2\; (2L-3)!\; (N-1)!^2 \;(L-N)!^2 }
{ (L-1)!^2 \; (2N-1)! \; (2L-2N-1)! } \, .
 \end{eqnarray}
  When   $ L \to \infty$,
  with a fixed  density $\rho = L/N$ and $ |j - L\rho ( 1 -\rho) |  \ll L$,   the 
  large deviation function $G(j)$, defined in~\eqref{def:LDF}   can be written in the 
 following scaling form: 
\begin{equation}
 G(j) =     \sqrt{ \frac{\rho ( 1 -\rho)} {\pi N^3}}
 H \Big( \frac{j -  L\rho ( 1 -\rho)}{\rho ( 1 -\rho)}   \Big) 
 \label{eq:scalf}
  \end{equation}
with 
\begin{eqnarray}
   H(y) \simeq  - \frac{ 2 \sqrt{3}}{5 \sqrt{\pi}} y ^{5/2} \,\,\,\,
 &\hbox{ for }&  \,\,\,\,   y \to +\infty \, , \\
     H(y) \simeq -  \frac{ 4 \sqrt{\pi}}{3} |y| ^{3/2} \,\,\,\,
 &\hbox{ for }&  \,\,\,\,   y \to -\infty \, . 
\end{eqnarray}
 This   large deviation function is not a quadratic polynomial, even in the vicinity
 of the steady state.  Moreover, the shape of this
 function is  skew: it decays as the exponential of a power law
 with an exponent $5/2$ for  $y \to +\infty$ 
 and with an exponent $3/2$ for  $y \to -\infty$.

\subsection{The general case: Functional Bethe Ansatz}

  In the general case $x\neq0$, the  Bethe Ansatz equations do not
  decouple  and a procedure for solving them was lacking.
  For example, it did not even seem  possible to extract from the
  Bethe equations~\eqref{eq:BAgamma} a formula for the mean stationary current
  (which can be obtained very easily by other means from the fact  that the stationary measure
  is uniform). Finally, the solution was found  by rewriting  the  Bethe Ansatz as a functional
  equation and  restating it as  a purely algebraic problem. We  explain here the method we followed
 \cite{Sylvain1,Sylvain3} and describe some of the results  obtained. \hfill\break
 First, we perform the following    change of variables, 
\begin{equation}
  y_i =  \frac{ 1 - {\rm e}^{-\gamma} z_i} { 1 - x {\rm e}^{-\gamma} z_i} \, .
  \end{equation}
  In terms of the variables $y_i$ the Bethe equations read
\begin{equation}
     {\rm e}^{L\gamma} \left( \frac{ 1-y_i}{1- x y_i} \right)^L =
      - \prod_{j =1}^N \frac{ y_i -  x y_j }{x y_i - y_j }
 \,\,\,\,   {\rm  for }\,\,\,\,  i =1 \ldots N\, .  
\label{eq:BAyi}
 \end{equation}
  Here again the  equations do not decouple as soon as $x \neq 0$. However,
 these equations are now built from first order monomials in the $y_i$'s
 and they are symmetrical in  these variables. This observation
 suggests to introduce an {\it  auxiliary variable} ${\bf T}$ that plays
 the same role with respect to all the  $y_i$'s and allows to define
 the auxiliary  equation: 
\begin{equation}
{\rm e}^{L\gamma} \left( \frac{ 1-{\bf T} }{1- x{\bf T} } \right)^L =
      - \prod_{j =1}^N \frac{ {\bf T} -  x y_j }{x {\bf T}  - y_j }
 \,\,\,\,   {\rm  for }\,\,\,\,  i =1 \ldots N\, . 
\label{eq:BAT}
 \end{equation}
 This equation, in which  ${\bf T}$ is the unknown, and the  $y_i$'s 
 are parameters, can be rewritten as a polynomial equation: 
\begin{equation}
   P(T) = {\rm e}^{L\gamma} (1 -   T)^L {\prod_{i =1}^N (x T - y_i)}
 + (1 - x T)^L  { \prod_{i =1}^N  (T -  x y_i)} = 0 \, .
 \end{equation}
  Because  the Bethe equations~\eqref{eq:BAyi} imply that  $P(y_i) = 0$ for
  $i =1 \ldots N\,$,   the polynomial $Q(T)$, defined as
 \begin{equation}
Q(T)=\prod\limits_{i=1}^{N}(T-y_{i}) \, , 
 \end{equation}
 must divide  the polynomial $P(T)$.  Now, if we examine  closely  the expression
 of  $P(T)$, we observe that the factors  that contain the $y_i$'s
 inside the  products over $i$  can be written  in terms of  $Q(T)$.
 Therefore, we conclude that    $Q(T)$  DIVIDES 
  $e^{L\gamma}(1-T)^{L}Q(xT)+(1-xT)^{L}x^{N}Q(T/x).$
 Equivalently, there  exists a   polynomial   ${R(T)}$    such that
\begin{equation}
 Q(T)R(T)=e^{L\gamma}(1-T)^{L}Q(xT)
  +x^{N}(1-xT)^{L}Q(T/x)  \,.
\label{eq:FBA}
 \end{equation}
 This functional equation is equivalent to the Bethe Ansatz equations
  (it is also known as Baxter's TQ equation). It can be used to determine
   the polynomial   ${Q(T)}$ of degree ${N}$  that  vanishes at the Bethe roots.
 In the present case, equation~\eqref{eq:FBA}
 can be  solved perturbatively  w.r.t. ${\gamma}$  to any desired order.
  Knowing   ${Q(T)}$  perturbatively  an expansion of 
   ${E(\gamma)}$ is derived, leading to the cumulants  of the current
 and to the  large deviation function. 
\hfill\break
 For example, this method allows to calculate the following cumulants
 of the total current:

    ${\bullet}$  {\it Mean Current $J$:}
    $ J=(1-x)\frac{N(L-N)}{L-1}\sim(1-x)L\rho(1-\rho)  \, \hbox{ for } \,\,
L\rightarrow\infty \, . $  \hfill\break

     ${\bullet}$  {\it Diffusion Constant $D$:} 
$${ D=(1-x)\frac{2L}{L-1}\sum_{k>0}{k^{2}\frac{C_{L}^{N+k}}{C_{L}^{N}}
\frac{C_{L}^{N-k}}{C_{L}^{N}}\left(\frac{1+x^{k}}{1-x^{k}}\right)} } \, . $$
 (We note that this  formula  was   previously  derived  \cite{DMal}  using
 the Matrix Product Representation that we discuss in the next section).
In the limit of a large system size, ${L\rightarrow\infty}$, with  asymmetry
 parameter  ${x\rightarrow 1}$ and  with a  {\it fixed value}  of
 ${ \phi=\frac{(1-x)\sqrt{L\rho(1-\rho)}}{2}}$, the diffusion constant assumes
 a simple expression
$$ {D\sim  4\phi L\rho(1-\rho)\int_{0}^{\infty}du\frac{u^{2}}{\tanh{\phi u}}e^{-u^{2}} }.$$

  ${\bullet}$   {\it  Third cumulant:}  the  Skewness measures the non-Gaussian
 character of the fluctuations.  An exact combinatorial expression of the third
 moment,  valid for any values
 of $L$, $N$ and $x$,   was calculated by S. Prolhac in \cite{Sylvain2}. It is given by 
\begin{eqnarray}
\frac{E_{3}}{6L^{2}}
&=&\frac{1-x}{L-1}\sum_{i>0}\sum_{j>0}\frac{C_{L}^{N+i}C_{L}^{N-i}C_{L}^{N+j}C_{L}^{N-j}}
{(C_{L}^{N})^{4}}(i^{2}+j^{2})\frac{1+x^{i}}{1-x^{i}}\frac{1+x^{j}}{1-x^{j}} 
    \nonumber \\
&-&\frac{1-x}{L-1}\sum_{i>0}\sum_{j>0}
 \frac{C_{L}^{N+i}C_{L}^{N+j}C_{L}^{N-i-j}}
{(C_{L}^{N})^{3}}
\frac{i^{2}+ij+j^{2}}{2}\frac{1+x^{i}}{1-x^{i}}\frac{1+x^{j}}{1-x^{j}} 
 \nonumber  \\
&-&\frac{1-x}{L-1}\sum_{i>0}\sum_{j>0}\frac{C_{L}^{N-i}C_{L}^{N-j}C_{L}^{N+i+j}}{(C_{L}^{N})^{3}}
\frac{i^{2}+ij+j^{2}}{2}\frac{1+x^{i}}{1-x^{i}}\frac{1+x^{j}}{1-x^{j}}
  \nonumber  \\
&-& \frac{1-x}{L-1}\sum_{i>0}\frac{C_{L}^{N+i}C_{L}^{N-i}}
{(C_{L}^{N})^{2}} \frac{i^{2}}{2}\left(\frac{1+x^{i}}{1-x^{i}}\right)^{2} 
  \nonumber  \\
&+& (1-x)\frac{N(L-N)}{4(L-1)(2L-1)}\frac{C_{2L}^{2N}}{(C_{L}^{N})^{2}} \nonumber  \\
 &-&(1-x)\frac{N(L-N)}{6(L-1)(3L-1)}\frac{C_{3L}^{3N}}{(C_{L}^{N})^{3}}  \, . 
\nonumber \end{eqnarray}
     For ${L\rightarrow\infty}$,   ${x\rightarrow 1}$   and keeping 
 ${ \phi=\frac{(1-x)\sqrt{L\rho(1-\rho)}}{2}}$ fixed, this formula  becomes
\begin{eqnarray}
&& \frac{E_{3}}{ \phi(\rho(1-\rho))^{3/2}L^{5/2}} 
  \simeq  -\frac{4\pi}{3\sqrt{3}} + 
  \nonumber  \\ 
&& 12  \int_{0}^{\infty}dudv
\frac{(u^{2}+v^{2}) e^{-u^{2}-v^{2}}-(u^{2}+uv+v^{2})e^{-u^{2}-uv-v^{2}}}
 {\tanh{\phi u}\tanh{\phi v}}   \, .
\nonumber \end{eqnarray} 
  This shows that the  fluctuations display  a  non-Gaussian behaviour.
 We remark that for   $\phi \to \infty$ the  TASEP limit is recovered:
 $$  E_{3} \simeq 
 \left(  \frac{3}{2} - \frac{8}{3\sqrt{3}} \right)
  \pi (\rho(1-\rho))^{2}L^{3} \, . $$
 
\hfill\break 

 A systematic expansion procedure that completely   solves the problem to all orders 
 and  yields exact expressions for all
 the cumulants of the current,   for an arbitrary value of the asymmetry parameter $x$,
 was carried out  by S. Prolhac in \cite{Sylvain4}.  Using  the functional Bethe Ansatz,
 S. Prolhac  derived  a  parametric representation
 of the cumulant generating function $E(\gamma)$ similar to the
 one given  for the TASEP in equations~\eqref{eq:EofY} and~\eqref{eq:gamY}, but where
  the binomial  coefficients are replaced  by combinatorial expressions that are  related
 to some tree structures.  A closed expansion of  $E(\gamma)$  w.r.t. $\gamma$ was
  derived and the  coefficients  that appear at each order were given 
  a  combinatorial  interpretation. This  expansion, valid for any finite values of $L$, $N$
 and $x$, was then used to study the large system size limit with various scalings
 of the asymmetry. Various  regimes were found and the corresponding expressions for the cumulants
  were fully worked out: 
\begin{itemize}
 \item  For  $ 1 - x \ll \frac{1}{L}$,  the model falls into the Edward-Wilkinson universality class.
 \item   The  range $ 1 - x  \sim  \frac{\nu}{L}$, where $\nu$ is a finite number, defines
 the weakly asymmetric regime (to be discussed below). 
  \item The intermediate  regime,  
 corresponding  to  $\frac{1}{L} \ll   1 - x  \ll   \frac{1}{\sqrt{L}}$,  exhibits 
 a specific scaling behaviour that,  to our knowledge, can not be represented by a continuous
 stochastic equation.
  \item  For
  $ 1 - x \sim  \frac{\Phi}{\sqrt{\rho(1-\rho)L}}$ the system is in  the strongly asymmetric regime.
 \item  Finally, $ 1 - x \gg  \frac{1}{\sqrt{L}}$ corresponds to the KPZ universality class,
 which contains  the TASEP. This limit was also studied by Lee and Kim \cite{LeeKim}.
\end{itemize}

\hfill\break

\begin{figure}[ht]
\begin{center}
   \makebox{
     \rotatebox{0} {
\begin{tabular}{ccc}
    \begin{tabular}{c}
      \includegraphics[height=3.75cm,angle =0]{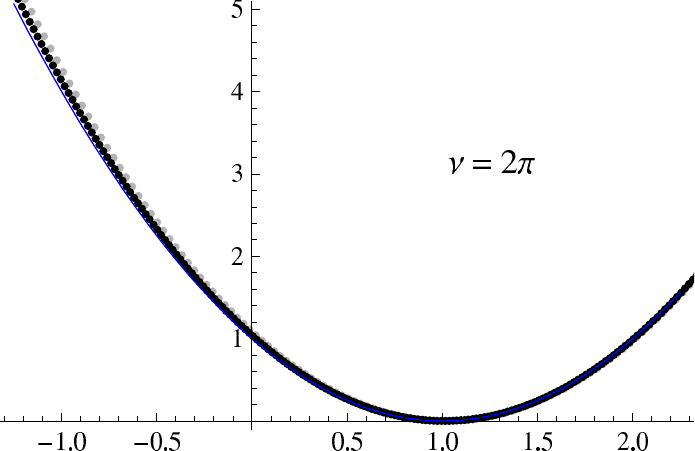} \\
       \includegraphics[height=3.75cm,angle =0]{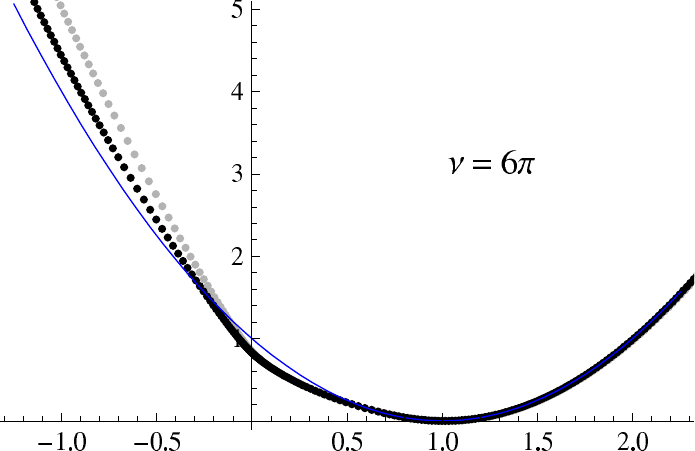} 
    \end{tabular} 
    & \quad  &
    \begin{tabular}{c}
   \includegraphics[height=3.75cm,angle =0]{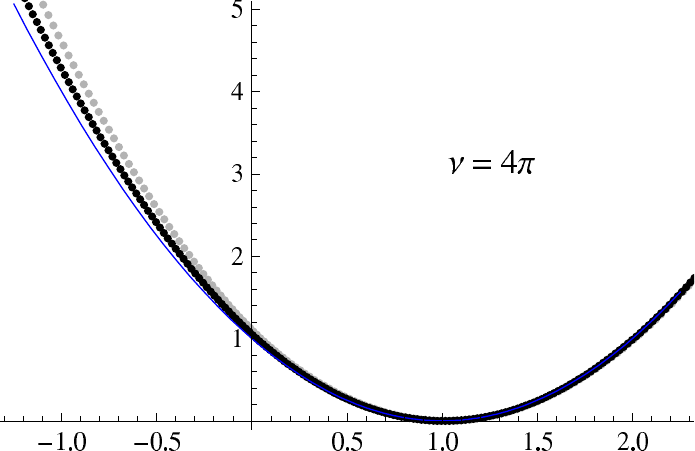} \\
         \includegraphics[height=3.75cm,angle =0]{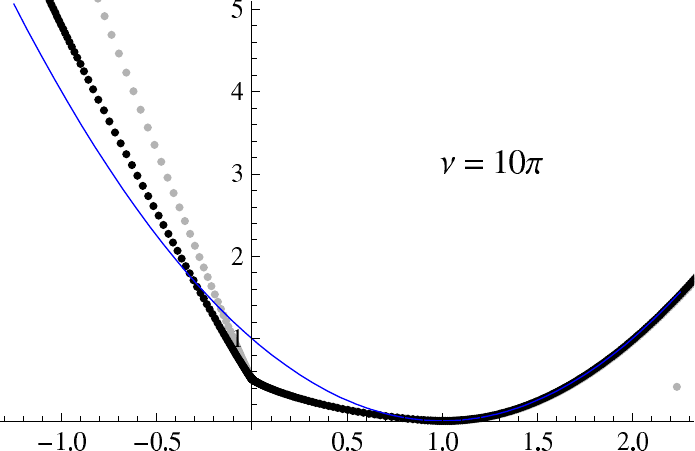} 
    \end{tabular} 
\end{tabular} 
 } } 
\end{center}
\caption{Behaviour of the large deviation function as a function
 of the current  $j/(\nu\rho(1-\rho))$  for different values of $\nu$.
  The gray dots correspond to  $L=50, N=25$ and the  black dots
 correspond to  $L=100, N=50$. They are obtained by solving numerically
 the functional Bethe Ansatz equation \eqref{eq:FBA}. The thin  blue curve
 represents the leading Gaussian behaviour~\eqref{LDFGaussian}.}
 \label{Fig:LDF}
 \end{figure}

 We conclude this section by some
 remarks specific to the  weakly  asymmetric case, for which the
 asymmetry parameter scales as  ${x=1-\frac{\nu}{L}}$ 
 in the  limit of  large system sizes  ${L\rightarrow\infty}$. 
 In this  case, we also need to rescale the fugacity parameter
 as ${\gamma}/{L}$ and 
 the  following  asymptotic formula for   the cumulant generating function
 can be derived 
 \begin{eqnarray}
  \tilde{E}(\gamma,\nu) \equiv
   E\left(\frac{\gamma}{L},  1- \frac{\nu}{L} \right)
 &\simeq&  \frac{\rho(1-\rho) (\gamma^{2} + \gamma\nu)}{L}
     -\frac{\rho(1-\rho)\gamma^{2}\nu}{2L^{2}}
  + \frac{1}{L^{2}} \phi [ \rho(1-\rho)(\gamma^{2} + \gamma\nu)] \, ,
 \label{eq:series1WASEP} \\
       \hbox{ with }   \,\,\,\,\,\,
   \phi(z) &=& \sum_{k=1}^{\infty} \frac{B_{2k-2}}{k!(k-1)!}z^k  \, ,
 \label{eq:series2WASEP}
 \end{eqnarray}
 and  where the ${B_{j}}$'s  are  Bernoulli Numbers. We observe
 that the leading order (in  ${1/L}$) is quadratic in $\gamma$
 and describes Gaussian  fluctuations. It is only in the subleading
 correction (in $1/L^{2}$) that the  non-Gaussian character arises.
 This formula was also obtained for the symmetric exclusion case
 $\nu = 0$ in \cite{Appert}.  We observe that the series
 that defines the function  $\phi(z)$ has a finite radius of convergence
 and that  $\phi(z)$ has a singularity for $z = -\pi^2$.  Thus,
 non-analyticities appear in $\tilde{E}(\gamma,\nu)$ as soon as 
   $$ \nu \ge \nu_c = \frac{2\pi }{\sqrt{\rho(1-\rho)}} \, . $$
  By  Legendre transform,  non-analyticities  also occur 
 in the large deviation function $G(j)$  defined in~\eqref{def:LDF}.
 At half-filling, the singularity appears at $\nu_c = {4\pi }$ as can be 
 seen in   Figure~\ref{Fig:LDF}.  For  $\nu < \nu_c$ the leading
 behaviour of  $G(j)$ is quadratic (corresponding
 to Gaussian fluctuations)  and is given by
 \begin{equation}
 G(j) = \frac{( j -\nu \rho(1-\rho))^2}{ 4 L \rho(1-\rho) } \, .
\label{LDFGaussian}
 \end{equation}
 For  $\nu > \nu_c$, the series expansions~\eqref{eq:series1WASEP}
 and  ~\eqref{eq:series2WASEP}  break  down  and 
 the  large deviation function $G(j)$  becomes
 non-quadratic even at leading order. This phase transition
 was predicted by  T.  Bodineau and B. Derrida using
 macroscopic fluctuation theory \cite{Bodineau1,Bodineau2}.
 These authors studied the optimal density profile that corresponds
 to the observation of the  total current $j$ over a large time $t$.
 They found that this optimal  profile is flat for  $j < j_c$ 
 but it  becomes linearly unstable for  $j > j_c$. In fact,
 when  $j > j_c$ the optimal  profile is  time-dependent. 
 The critical value of the total current  for which this phase-transition occurs
 is  $j_c =\rho(1-\rho) \sqrt{ \nu^2 -  \nu_c^2}$ and therefore
 one must  have  $\nu \ge  \nu_c$ for  this transition to occur. 
 One can observe in  Figure~\ref{Fig:LDF} that for $\nu \ge  \nu_c$,
 the   large deviation function  $G(j)$  becomes non-quadratic and develops
 a kink at a  special value of the total current $j$. 

 \section{Multispecies exclusion processes and Matrix Ansatz}
\label{sec:Multi}

 From the mathematical point of view, the ASEP is one of the simplest
 but non-trivial model for which the hydrodynamic limit can be rigorously
 proved. At large scales, the  distribution of the
  particles of the ASEP emerges as  a density field that
 evolves according to Burgers equation with a vanishingly small viscosity \cite{Spohn}.
 The Burgers equation is the textbook  prototype for shock formation: 
 a smooth initial distribution can develop a singularity (a discontinuity)
 in finite time. A  natural question that arises  is whether  this   shock 
 is an artifact of the hydrodynamic limit or whether, under some
 specific conditions,  the original ASEP does display some singularity
 at the microscopic scale \cite{Andjel,Ferrari,Barma2,MartinSchock,Janowski,km}. 
  The question was answered positively:
  an abrupt transition does exist  at the  level of the particle system
  and its   width  is of the order of the
 lattice size. However,  defining  precisely the  position of the 
 shock at  the microscopic level  requires some thought, and this was achieved
 by introducing a second-class particle (see e.g., P. A. Ferrari, 1992 \cite{Ferrari}). 
  This second class particle behaves as a first-class  particle with respect to holes
 and as a hole with respect to first-class  particles.  The dynamical rules thus take the
  following  form:
 $10 \to 01  \,\,\,\,  12 \to 21 \,\,\,\, \hbox{ and } \,\,\,  20 \to 02 \,,$ 
 where all transitions occur with rate 1.
 This dynamics corresponds to  coupling  two TASEP models. 
 In order to  locate the  shock it is enough
 to introduce  a single second class particle in the system.

A straightforward  generalization of the two species case of first and second
class particles is  the multispecies process where there is a
hierarchy amongst the different species: first class particles have highest
 priority and can overtake all other classes of
  particles; second class  particles can overtake  all other classes
 except the first class etc...
 hence,  during an infinitesimal time step $dt$, the following processes
  take place on each  bond  with probability $dt$:
\begin{eqnarray}
 I\ 0 \to 0\ I&&\qquad \mbox{for}\quad I \neq 0  \nonumber\\
I\ J  \to J \ I &&\qquad \mbox{for}\quad 1 \leq I <  J \leq N \, .
  \label{eq:dynRules}
\end{eqnarray}
Note that   particles can always overtake holes (=  0-th class particles).
  This  model  will be called  the $N$-TASEP and   it can be obtained by coupling
   $N$  ordinary  TASEP models \cite{Liggett1,Liggett2}.

\hfill\break
 Suppose that  there are $P_I$ particles of class  $I$ on a ring of size $L$. 
 Then,  the total number of configurations is given by 
$$ \Omega = \frac {L!}{  P_0! P_1!P_2!\ldots P_N! }$$
The total number of particles is given $P_1 + \ldots  P_N$. We warn the reader that in this
 section $N$ denotes the  number of species and  not the number of particles. 
   
  The object of this section
 is to provide an algebraic description
 of the   stationary measure of this system.

\subsection{Matrix Ansatz for Two Species}

     The idea of representing the stationary  weights of a  configuration
 as traces over a quadratic algebra goes back to the seminal article
 of Derrida, Evans, Hakim and Pasquier in 1993 \cite{DEHP}.  These authors
 were studying the 1-species TASEP with open boundary conditions
 but they   realized that the same idea  could also be  applied  to the 
  Two Species ASEP on a ring (Derrida, Janowski, Lebowitz
 and Speer, 1993  \cite{DJLS}). 
 This technique, known as the Matrix Product
  Representation (or Matrix Ansatz), 
 has been exceptionally fruitful in the field of one-dimensional
 stochastic models and has led to a very large number of exact solutions.
 This method seems to be complementary to the Bethe Ansatz in many
 problems and the exact relations between the two techniques
 is not yet fully understood \cite{Alcaraz2,ogkmMP}. 
  A solver's guide to  the  Matrix Ansatz method
  has recently been written by R.A.  Blythe and M. R. Evans
 \cite{MartinRev2}  and contains many applications to various models.

  Here, we explain how the Matrix Ansatz works for the  two species
 TASEP on a ring, with  dynamical rules given by~\eqref{eq:dynRules}.
 A Configuration of the model can be
   represented by  a string made from  the 'letters' 1,2 and 0,  e.g., 
  ${\mathcal C} = {0}{1}{22}{0}{2}{11}$. According to the  Matrix Ansatz, 
  the   stationary weight  of    ${\mathcal C}$  is  given by 
\begin{equation}
 P({\mathcal C}) =  P({0}{1}{22}{0}{2}{11}) =
 \frac{1}{Z}{\rm Tr}( {E}{D}{AA}{E}{A}{DD} ) \, ,
\label{MatrixAnsatz}
 \end{equation}
 where the string corresponding to   ${\mathcal C}$  has been rewritten
  using the alphabet $D,A$ and $E$ through the correspondence: 
 ${0}~\rightarrow~{E},$  ${1}~\rightarrow~{D}$
  and ${2}~\rightarrow~{A}$. Here  $D,A$ and $E$  are operators
  (matrices) which are in general non-commuting;
  we suppose that the Trace operation is well-defined
 on any product that contains at least one operator of each type.
 If the particles were totally independent 
  then the   stationary measure would be factorized
 and the probability of having a 0,1 or 2 
  at  a site will be equal, respectively, 
  to the density $\rho_0$ of holes,  $\rho_1$ of 1's or  $\rho_2$ of 2's.
 Of course, this is not true:
  the particles are strongly correlated by the dynamics. 
 The Matrix Product Representation can  be  `justified' a posteriori 
  in an informal way by saying  
 that in  this  Ansatz   
  the  stationary measure  remains somehow factorized 
  but correlations  are  taken into account    because the
 operators  $D,A$ and $E$  do not commute. As usual, the factor $Z$
 in equation~\eqref{MatrixAnsatz} is a normalization factor. 

  The  operators $D,A$ and $E$  have to be  chosen adequately 
 so that the Ansatz~\eqref{MatrixAnsatz} corresponds to the
 zero-eigenvector of the Markov matrix. It can be shown that
 the right choice in the case of the 2-TASEP model on a ring is
 obtained when these operators satisfy the following relations:
    \begin{eqnarray}
  {D}{E} &=& {D} + {E} \,,   \label{eq:algebre1}  \\
 {D}{A}  &=& {A}    \,,  \label{eq:algebre2}   \\
    {A} {E} &=& {A} \label{eq:algebre3}  \,. 
 \end{eqnarray}  
   The operators  $D,A$ and $E$ thus define a  {\it quadratic algebra}.
 These algebraic rules are enough to allow the computation of the
 steady state probability of any configuration. For example, 
 $$ P({0}{1}{22}{0}{2}{11}) = \frac{1}{Z}
{\rm Tr}(D^2EA^3) = \frac{1}{Z} {\rm Tr}((D^2 + D +E)A^3)  
 =   3  \frac{1}{Z} {\rm Tr}(A^3). $$
 The overall normalization constant can also  be determined. 
  It is possible to prove from 
  algebraic 
 relations~(\ref{eq:algebre1}, \ref{eq:algebre2}, \ref{eq:algebre3})
 that the weights defined by the  Matrix Ansatz  correspond
 to the stationary probabilities. This  Ansatz allows to calculate
 many  stationary state properties such as 
 currents, correlations, fluctuations \cite{DJLS}.  Besides, 
 it can be shown, using  Matrix Ansatz, that the stationary measure 
 of the Two Species  TASEP on a ring is not Gibbsean 
(Speer, 1993 \cite{Speer}).

 The algebra~(\ref{eq:algebre1}, \ref{eq:algebre2}, \ref{eq:algebre3})
  encodes combinatorial recursion relations between  configurations
 corresponding to different system sizes.  Although most of the
 calculations can be done  just by using  the abstract algebraic
 relations between $D$, $E$ and $A$ that define the algebra,
  it can be sometimes helpful to work with an explicit representation.
   The non-commuting representations of the 
  algebra~(\ref{eq:algebre1}, \ref{eq:algebre2}, \ref{eq:algebre3})
 are necessarily  infinite dimensional. One of the most 
popular representation is the following: 
\begin{eqnarray}
  D  = \left( \begin{array}{ccccc}
                  1&1&0&0& \dots\\
                  0&1&1&0& \\
                  0&0&1&1&\ddots\\
                  &&&\ddots&\ddots 
                   \end{array}
                   \right), \,\, 
  E  =  D^{\dagger},  \,\, 
   A  = \left( \begin{array}{cccc}
                  1&0&0&\dots\\
                   0&0&0&\dots\\
                 0&0&0& \dots \\
                 .&.&.&.
                   \end{array}
                    \right) \,\; \nonumber 
 \end{eqnarray}
 These matrices operate on an infinite dimensional space
 with countable basis.
It we write 
 $ D = 1 + \delta$ and $ E = 1 + \epsilon $, we note that 
  ${\delta}$ corresponds  the right-shift and $\epsilon$ to 
  the left-shift.
 The operator $A$   is simply  the  projector on  first coordinate.
 We also remark that any finite product of the matrices  $D$, $E$ and $A$
 that contains at least one $A$ is a finite rank matrix and thus has  a finite
 trace. 
 
    Apart from an isolated attempt to solve the 3-TASEP 
 using the Matrix Ansatz that did not seem to extend  to higher class
 models \cite{MMR}, no representation for the  $N-$TASEP  stationary measure
 was known until 2007. The breakthrough was made in the mathematical literature
 and came from a very different
 direction,  queuing theory.

\subsection{Geometric interpretation of the  2-TASEP stationary measure}

 In~\cite{FerrariMartin}, 
 P. A. Ferrari and J. B. Martin  found  an independent construction
 of the   2-TASEP steady state that does not rely on the  Matrix Ansatz
 but rather on a  queuing model interpretation.  These authors
 construct a  2-TASEP configuration with $P_1$  First  Class Particles
 and $P_2$ Second Class Particles  starting
  from two independent configurations of the 1 species TASEP defined
 on two  parallel  lines. 

\begin{figure}[ht]
\begin{center}
\begin{tabular}{c}
   \includegraphics[height=2.0cm]{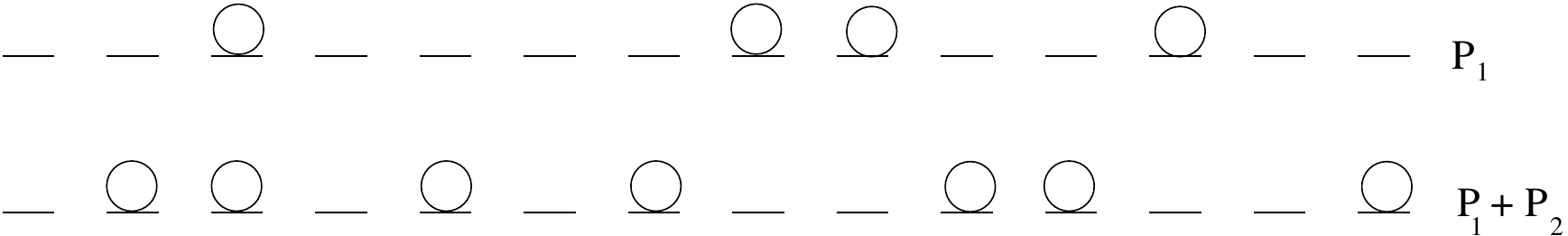} \\ \\  \\ \\ \\
   \includegraphics[height=2.0cm]{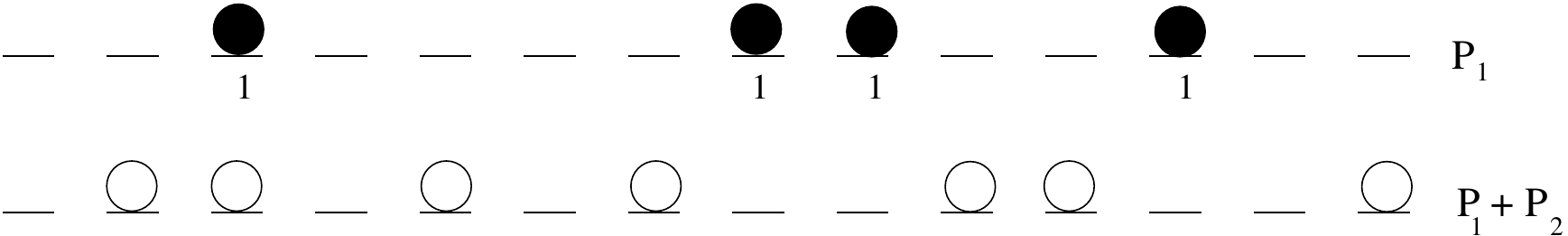}  \\ \\   \\ \\ \\
    \includegraphics[height=2.0cm]{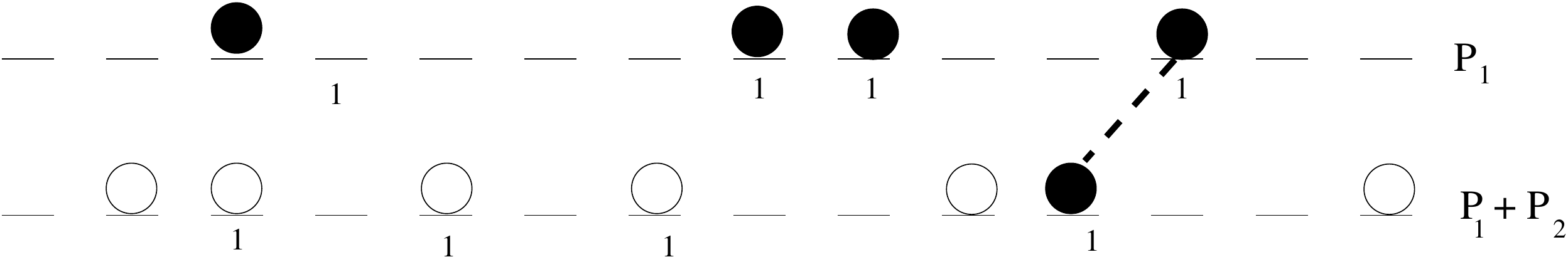} \\ \\  \\ \\ \\
  \includegraphics[height=2.0cm]{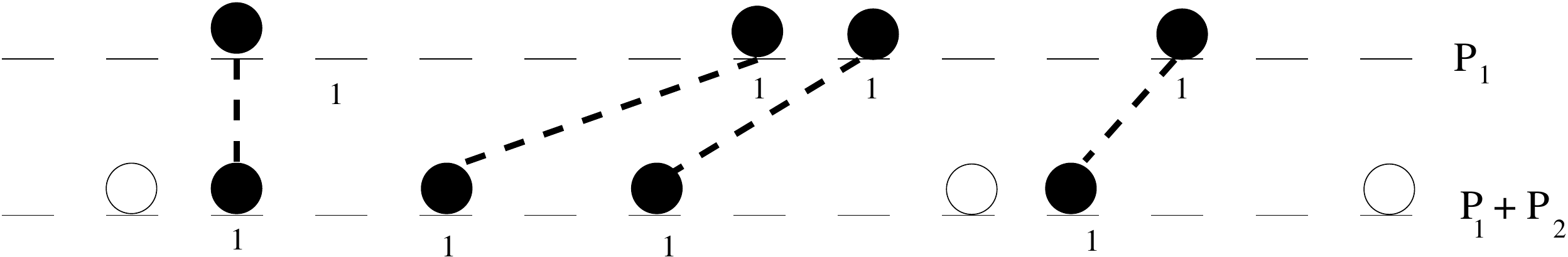} \\ \\   \\ \\ \\
   \includegraphics[height=2.0cm]{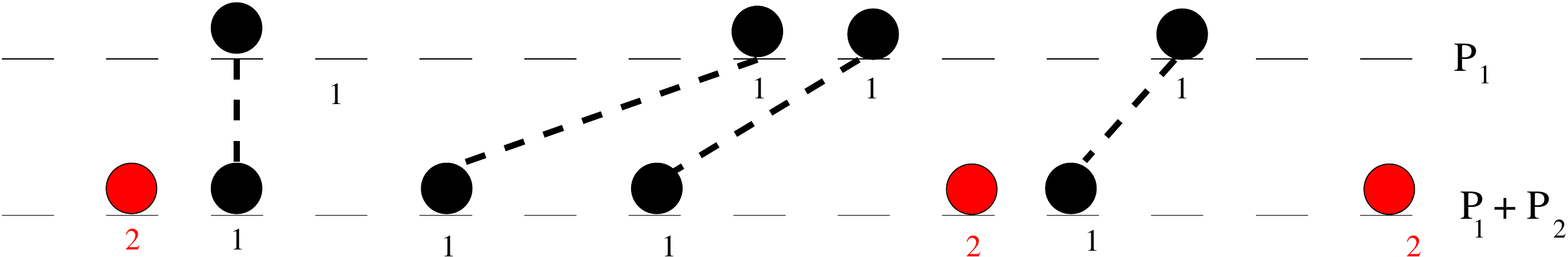}
 \end{tabular} 
 \caption{Step by step decomposition  of the Ferrari and Martin construction.}
\label{fig:FerrariMartin}
\end{center}
\end{figure}

\begin{figure}[ht]
\begin{center}
   \includegraphics[height=5.0cm]{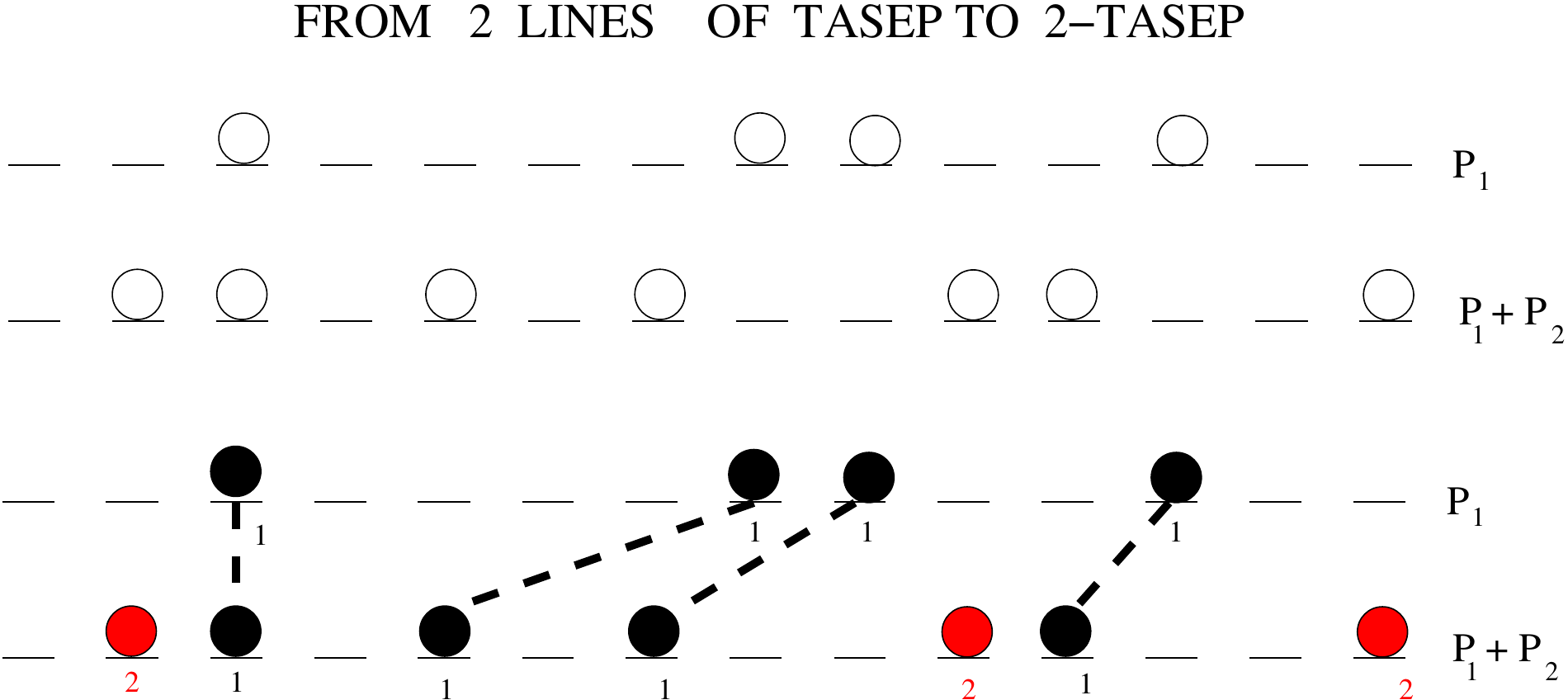}
  \caption{A synthetic  view of the Ferrari and Martin construction:
 starting from two lines of TASEP with only one type 
 of (colorless) particles, a 2-TASEP configuration is constructed
 with first class particles (in black), second class particles (in red)
 and holes (empty sites).}
\label{fig:FMresume} 
 \end{center}
\end{figure}

 One starts from  a two-line
configuration of particles (with at most one particle per site).
 The following algorithm   will be  easier to follow  by
 looking at    Figure~\ref{fig:FerrariMartin}  in which the construction
 in drawn step by step:
   On line 1 there are $P_1$
particles distributed randomly  amongst the $L$ available sites.  On line
2,  there are $P_1+P_2$ particles also  distributed randomly amongst the $L$  sites. 
  Working from right to
left we associate to each particle on line 1, the nearest particle, at the 
site just below it or  to its  left, on line 2 that has not been already associated to another particle.
The  particles in line 2 that are paired to particles on the first line
  are  labeled 1;  the  $P_2$  particles  on line 2  that remain 
unassociated  are labeled 2. The empty sites of line 2 are labeled 0, 
thus each site of line 2 is labeled 0, 1 or 2.  Hence,  a configuration of
the two species TASEP containing $P_1$ first-class and $P_2$ second-class
particles is  generated by  the construction.
 Since we consider
periodic boundary conditions, the site at which we begin this procedure 
 will  not affect the
two species TASEP configuration that is obtained.
 The procedure is  summarized in Figure~\ref{fig:FMresume}.

\hfill\break

 It is
 important to note that this  construction
 is NOT one-to one: different configurations of the $P_1$ particles  on line 1 can lead
 to the same final 2-TASEP configuration on line 2. 
 The fundamental claim, proved in~\cite{FerrariMartin},  is the following: 
{\it  the weight of a 2-TASEP configuration is proportional
 to the total  number of ways you can generate it by the 
Ferrari-Martin construction.}
  By examining carefully  the Figure~\ref{fig:FMresume},
  one can make the following  fundamental  observations:
  \begin{itemize}
 \item[$\bullet$]  A  particle 1 (on the 1st line)  can not be located above a 2
 on the 2nd line. 
  \item[$\bullet$]  { {\bf Factorisation Property:}}
 All  the  1's (on the 2nd line) situated  between
 two 2's MUST be linked to   1's (on the 1st line) that  are   located
 between the positions of the  two 2's {\it (No Crossing Condition)}.
 \item[$\bullet$]  { {\bf `Pushing'  Procedure:}}
  The {\it  `ancestors'} of   a string of the
 type $210102$  are  the  strings obtained by pushing the 1's
 to the right i.e., $210102,210012,201102,201012,200112$.  
 \end{itemize}

  In fact, these properties uniquely characterize the stationary  weights.
  Furthermore, one can prove that the matrix Ansatz automatically
 performs the combinatorics underlying the geometrical  construction
 of the weights.  More precisely, one can construct in a systematic manner 
 the quadratic algebra
 generated from  $A,D$ and $E$ from  the Ferrari-Martin procedure \cite{EFM}; 
  reciprocally one can deduce this  procedure from the
  algebra~(\ref{eq:algebre1}, \ref{eq:algebre2}, \ref{eq:algebre3}).
 This  correspondence leads to the following properties:
\begin{itemize}
  \item[$\bullet$]  {{\bf The  Factorisation Property}} 
 is related to the fact  that  $A$ is a  {{\bf PROJECTOR.}} 
\item[$\bullet$]  {{\bf The  Pushing Procedure}}  leads
 to the fact  that$D$ and $E$ are 
    {{\bf SHIFT OPERATORS}}  (right-shift
  and left-shift, respectively).
 \end{itemize}

\subsection{Solution of the   $N$-TASEP}

\begin{figure}[ht]
\begin{center}
  \includegraphics[height=3.0cm]{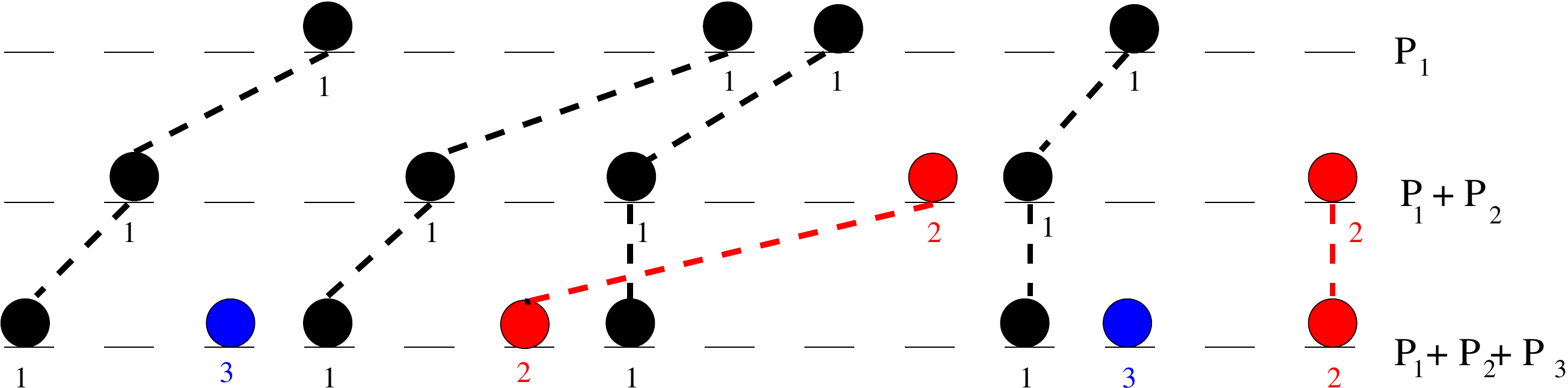}
  \caption{The Ferrari and Martin construction for
 the 3-TASEP . On the last line, 
  a 3-TASEP configuration is generated
 with first class particles (in black), second class particles (in red)
 third class particles (in blue)  and holes (empty sites).}
\label{fig:FM3TASEP} 
 \end{center}
\end{figure}

 The geometric construction of  Ferrari-Martin was carried out
 recursively to the $N$-TASEP model \cite{FerrariMartin}.
 In Figure~\ref{fig:FM3TASEP},  
  the  3-TASEP is described:  one
 starts  with 3 lines of the 1-TASEP, containing respectively
 $P_1$,  $P_1 + P_2$ and   $P_1 + P_2 +P_3$  particles with no labels.
 All the  $P_1$  particles on the first line are  in black and
 considered to be  1st-class particles. 
 One starts by  pairing  from right to left the 
 1st-class particles  from the 1st  line to the 2nd line.
  This procedure is repeated from the  2nd line  to the  3rd  line  on so on.
  Each line has thus $P_1$  1st-class particles,
  in black. These 1st-class particles are now considered
 to be spectators. 
The $P_2$  unselected particles  on the second line are 2nd-class  particles,
 they are colored in red. We associate the   $P_2$ red-particles
 of the  second line to  $P_2$ particles on the third line again
 from  right to left.  Now on the  third line, we have 
  $P_1$  1st-class particles (in black),   $P_2$   2nd-class particles (in red), and  
 the  remaining,  unselected,  $P_3$ particles  are  3rd-class
 particles (in blue). We have  thus constructed a  3-TASEP configuration
 on the third line.    This method  extends to the $N$-TASEP
  in a natural manner (for a precise description see 
 \cite{FerrariMartin,EFM}).

  Here again,  the same  3-TASEP configuration on the third  line 
 can be generated   in different ways.
 The  weight of a 3-TASEP configuration is proportional
 to the total  number of ways you can generate it by this construction.
  The Matrix Ansatz for the  3-TASEP  is the  algebraic counterpart
 of this procedure.
  It was shown  in \cite{EFM}  that  the recursive
  geometrical  construction leads to a   hierarchical definition of the
  operators  $\hat{\bf P}_0,\hat{\bf P}_1, \hat{\bf P}_2$ and 
 $\hat{\bf P}_3$
 corresponding to the particles $0,1,2,$ and 3 based
  on{\it  Tensor Products} of elements of the original  quadratic algebra.
  Indeed, for the 3-TASEP case, we find 
 using   $D$, $A$ and $E$ matrices and the shift operators
 $\delta = D -1$ and $\epsilon = E-1$:
\begin{eqnarray}
          \hat{\bf P}_0      &=&     {\bf 1}\otimes{\bf 1}\otimes E +
           {\bf 1}\otimes  \epsilon  \otimes A + 
   \epsilon  \otimes  {\bf 1}\otimes D \, .    \label{def:P0} \\ 
       \hat{\bf P}_1  &=&   {\bf 1}\otimes{\bf 1}\otimes D  + 
   \delta \otimes  \epsilon  \otimes A +   \delta \otimes {\bf 1}\otimes E
       \label{def:P1}  \\
     \hat{\bf P}_2 &=&  A  \otimes {\bf 1}\otimes A
        +    A  \otimes  \delta \otimes E         \label{def:P2}   \\
     \hat{\bf P}_3 &=&    A  \otimes A    \otimes E   \label{def:P3}
          \end{eqnarray}
 More generally, the    Ferrari-Martin construction allows to construct
 a Matrix Ansatz for the  $N$-TASEP and also provides
 an interpretation of the operators that appear in the 
 Matrix Ansatz as priority queue matrices~\cite{EFM}. 

 We now discuss the case of the $N$-ASEP model.
 If  backward jumps  are  allowed (rate $x \neq 0$), the $N$-ASEP
 dynamical rules are  
\begin{eqnarray}
JK\to KJ&\mbox{\quad with rate $1$ \quad if $1\leq J<K\leq N$}\\
KJ\to JK&\mbox{\quad with rate $x$ \quad if $1\leq J<K\leq N$}\\
J0\to 0J&\mbox{\quad with rate $1$ \quad if $1\leq J\leq N$}\\
0J\to J0&\mbox{\quad with rate $x$ \quad if $1\leq J\leq N$}\;.
\end{eqnarray}
  Unfortunately, the  Ferrari-Martin  construction  
{\it can not be generalized} to the $N$-ASEP model. This can be understood
 by the fact that the procedure  for the TASEP
  is local and directed (a particle
 is associated with the nearest particle  on its left in the next line)
 and by introducing backward jumps 
the queuing theory interpretation is lost.
 In fact it can be shown that no  construction  similar to that 
  of  Ferrari and Martin  can exist for the  $N$-ASEP and this method can  not be used
 to determine the  steady state measure of the $N$-ASEP on a ring.

  However,  using   the operators defined in
  equations~(\ref{def:P0}--\ref{def:P3}) and  deforming   the underlying
 quadratic algebra in the following manner:
\begin{eqnarray}
DE -x ED &=& (1-x)(D + E) \label{DE}\\
DA-x AD &=& (1-x) A\label{DA}\\
AE -x EA &=& (1-x)A \; ,  \label{AE}
\end{eqnarray}
  we proved  in  \cite{PEM} that the
  deformed tensor algebra  solves the  $N$-ASEP.
 We note  that the 
 shift-operators,  still defined by $ \delta =  D- 1$
 and $\epsilon = E - 1$, satisfy  the  {\it $x$-deformed  oscillator algebra}:
 $$ \delta\epsilon  -  x \epsilon \delta = 1 \, ,$$
  as can be shown  using  equation~\eqref{DE}.
  In fact, a similar  deformation method  allows to find  the steady state of the
   $N$-ASEP as proved   in  \cite{PEM}.

 To conclude, we emphasize that  in the present problem,
   the Matrix Ansatz is not simply  a reformulation 
  of a known  algorithm  but it also plays a constructive role and allows
 to derive  new results which are not accessible by other means. 
 Besides, in the TASEP case, the    Ferrari and Martin  construction
 provides for the first time  \cite{EFM} an interpretation  of the
 infinite-dimensional  operators  that appear in the  Matrix Ansatz: the mapping
 to a queueing process provides a natural representation of the space on which the
 matrices act as counting operators
 (for an alternative interpretation of the  Matrix Ansatz
 see \cite{Woelki}).  There are many ways to generalize
 the TASEP to multispecies models (by introducing e.g. 
  species-dependent switching rates, or open boundary conditions \cite{Ayyer})
 and we believe that  the  algebraic deformation technique explained above
 can  be adapted  to  various  unsolved problems. The $N$-ASEP also displays  
 remarkable properties that have been investigated  by combinatorial methods 
 \cite{Chikashi} and partially  by the Bethe Ansatz for $N=2$ 
 \cite{Cantini,evans,Wehefritz} but many open problems remain to be solved.

\section{Conclusion}

  The asymmetric  exclusion process seems deceptively simple and  
  the incredible number and complexity  of the studies that it stimulated 
  appears totally incredible.  Such  a  wonder  seems  to defy  explanation.
   Very  few  models have met such a rare success: one  
   example that comes to  mind is the Ising model.

  The present article stems from  a talk given at STATPHYS24 and discusses 
  some recent  works. It  does not pretend to be an exhaustive review.
  We  focused on models on a finite periodic lattice  that we analyzed by 
  two techniques,  the  Bethe Ansatz and the matrix product representation.
  The bibliography, although quite substantial, reflects this subjective choice.
  The  ASEP  can be investigated   through
  a  huge  variety of techniques:  Bethe Ansatz, Quadratic Algebras, 
  Young tableaux,  Combinatorics,  Orthogonal polynomials, 
  Random Matrices,   Stochastic differential equations, 
  determinantal representations, 
  hydrodynamic limits etc.
  Each of these approaches is becoming   a specific  subfield 
  that has its  own links with other branches of
  theoretical physics and  mathematics.
  We refer the reader who wishes
  to  broaden   his (her) perspective to 
  recent review articles   and to other contributions to this special issue.
  In particular,
   the ASEP on  the  infinite lattice,  the  relations  with
   random matrix theory
   and the recently derived exact results for the Kardar-Parisi-Zhang
  equation in one dimension, are reviewed  in 
  the articles  of P. L.  Ferrari and of 
  H. Spohn and T. Sasamoto.

  \vskip 0.5cm

   I am grateful to my  colleagues,  C. Arita, A. Ayyer, M. Evans,  P. Ferrari,
   O. Golinelli, S. Prolhac, N.~Rajewski  and M. Woelki,   with whom 
   I had the pleasure to work on the subjects presented
   here during the last years.  I thank  S.~Mallick for collaboration 
   and for his careful reading of the manuscript.

\end{document}